\documentstyle[12pt,apjpt4]{article}

\batchmode
\singlespace

\lefthead{Jernigan, Klein and Arons}
\righthead{Discovery of kHz Fluctuations in Centaurus X-3}

\begin{document}

\title{Discovery of kHz Fluctuations in Centaurus X-3:\\
Evidence for Photon Bubble Oscillations (PBO) \\
and Turbulence in a High Mass X-ray Binary Pulsar}

%\title{Discovery of kHz Quasi-Periodic Oscillations (QPO) \\
%and Continuum in Centaurus X-3:\\
%The Existence of Photon Bubble Oscillations (PBO) \\
%and Turbulence in a High Mass X-ray Binary Pulsar}

\author{{J. Garrett Jernigan}$^{1}$}
\author{{Richard I. Klein}$^{2,3,5}$}
\author{{Jonathan Arons}$^{2,4,5}$}
\received{ApJ: April 1, 1999}
\accepted{ApJ: September 1, 1999}

\altaffiltext{1} {Space Sciences Laboratory, University of California, Berkeley, CA 94720. \\
e-mail: jgj@xnet.ssl.berkeley.edu}

\altaffiltext{2} {Department of Astronomy, 601 Campbell Hall, Univ. of California, Berkeley, CA. 94720. \\
e-mail: klein@radhydro.berkeley.edu,
arons@astroplasma.berkeley.edu}

\altaffiltext{3} {Lawrence Livermore Nat'l Lab., Univ. of California L-23, P.O. Box 808, Livermore, CA. 94550}

%\altaffiltext{4} {Institute of Geophysics and Planetary Physics, Lawrence %Livermore Nat'l Lab. L-491, Univ. of California, P.O. Box 808, Livermore, CA 94550}

\altaffiltext{4} {Department of Physics, University of California, Berkeley, CA 94720}

\altaffiltext{5} {Theoretical Astrophysics Center, University of California, Berkeley, CA 94720}

\begin{abstract}

We report the discovery of kHz fluctuations, including quasi-periodic
oscillations (QPO) at $\sim330~$Hz ($260-407~$Hz)
and $\sim760~$Hz ($671-849~$Hz) and a
broadband kHz continuum in the power density spectrum
of the high mass X-ray binary pulsar Centaurus X-3 (\cite{jernigan99}).
These observations of
Cen X-3 were carried out with the Rossi X-ray Timing Explorer (RXTE).
The fluctuation spectrum
is flat from mHz to a few Hz, then steepens to  $f^{-2}$ behavior between
a few Hz and $\sim 100$ Hz. Above a hundred Hz, the spectrum shows the
QPO features, plus a flat continuum extending to $\sim1200~$Hz and then
falling out to $\sim1800~$Hz.
%Signal-to-noise limitations prevent the precise characterization
%of the spectrum to high frequencies ($1200-1800~$Hz).
These results, which required the co-adding three days of
observations of Cen X-3, are at least as fast as the fastest
known variations in X-ray emission from an accreting compact object
(kHz QPO in LMXB sources) and probably faster since extension to $\sim 1800~$ Hz is
indicated by the most likely parameterization of the data.

Multi-dimensional radiation hydrodynamics simulations of optically
thick plasma flow onto the magnetic poles of an accreting neutron star
show that the fluctuations at frequencies above 100 Hz are
% the likely consequence of the
consistent with
photon bubble turbulence and oscillations (PBO) previously
predicted (\cite{klein96a}) to be observable in this source. We show that
previous observations of Cen X-3 constrain the models to depend on
only one parameter, the size of the polar cap.  For a polar cap opening
angle of 0.25 radians (polar cap radius $\sim 2.5$ km and area $\sim 20$ km$^2$,
for a neutron star radius of 10 km), we show that the spectral form above
100 Hz is reproduced by the simulations, including the
frequencies of the QPO and the relative power in the QPO
and the kHz continuum. This has resulted
in the first
% indirect
model-dependent
measurement of the polar cap size of an X-ray pulsar.
%Definite quantitative measurement is limited by the finite number of
%theoretical numerical models and the restriction to 2D symmetry.
%However very large polar caps ($>0.4~$radians) or small polar caps
%($<0.1~$radians) are not consistent with the observed data.
The simulations underpredict the overall amplitude of the observed spectrum,
which we suggest is the consequence of a 2D axisymmetric
simulation of an intrinsically 3D phenomenon.
The power density spectrum of Cen X-3 shows a
dramatic decrease above $\sim1000~$Hz which suggests an optical depth
$\sim30$ across the accretion mound consistent with effects of radiative
diffusion in the simulations.
We identify this decline at high frequency as the first direct evidence
of radiative diffusion near the surface of a neutron star (NS).

We suggest the fluctuations observed at frequencies below 100 Hz,
whose spectrum has a different form from that of the kHz phenomena,
reflect intermittency in the mass transfer mechanism which carries
plasma from the accretion disk to field aligned flow onto the neutron star's
polar caps.  Using simple estimates based on Rayleigh-Taylor instabilities,
possibly modulated by intrinsic disk turbulence, we show that mass
transfer in ``blobs'' forming through Rayleigh-Taylor disruption of the
disk's inner edge can explain the large amplitude fluctuations required
by the spectrum at frequencies $f \sim 1$ Hz, but only if magnetic pressure
in the disk's innermost regions inflates the disk until its scale height
is comparable to the magnetosphere's size $\sim 4300$ km.

The observational results required the development of a procedure for
the careful determination of the deadtime effects of the PCA.
This procedure is described in appendix A.
As a consequence of the use of observations of Cyg X-1 for
the estimation of deadtime corrections for the observations of Cen X-3,
we have also demonstrated that the black hole Cyg X-1 shows clear
evidence of variability up to a frequency of $\sim280~$Hz.
Also observations of GX 17+2 were used to validate the procedure for deadtime
corrections. This analysis of GX 17+2 clearly
indicates the presence of a kHz QPO and the absence of any 
significant simultaneous kHz continuum.

\end{abstract}


\section{Introduction}

The binary system composed of the X-ray emitting neutron star Cen
X-3 and its companion O-type supergiant has been studied
extensively by all orbiting X-ray astronomy observatories. The
results from the Uhuru satellite identified the system as an
eclipsing binary X-ray pulsar with a 2.087 day orbit and a 4.84 s
pulse period (\cite{giacconi71}; ~\cite{schreier72}). The details
of the orbit determination, which include estimates of the masses
of both the neutron star and its companion, are summarized in a
review (\cite{rappaport83}).
The well determined distance to the binary system (\cite{hutchings79}) yields
an accurate determination of the luminosity.
Recently a cyclotron feature was detected in the X-ray spectrum of Cen X-3
providing precise knowledge of its surface magnetic field strength
(\cite{sax98}).

%Table~\ref{tblcenx3}
Table 1
summarizes the parameters of Cen X-3.
The average X-ray luminosity is
\begin{equation}
   L_{x}~=~
   9.4~\times~10^{37}
   {\left[ D \over 10~{\rm kpc} \right] }^2
   {\left[ {{(1 + Z)} \over {1.3}} \right] }^{2} \; {\rm erg~s^{-1}}
\label{lumin}
\end{equation}
where $Z = ((1-2GM/Rc^{2})^{-1/2} -1)$ is the surface gravitational redshift
of the star of mass $M$ and radius $R$.
The surface magnetic field strength $B$ implied by the observed cyclotron feature is
\begin{equation}
   B~=~
   3.2~\times~10^{12}
   {\left[ {E_{c} \over {28.5~{\rm keV}}} \right]}
   {\left[ {{(1 + Z)} \over {1.3}} \right] } {\rm Gauss}
\label{cyclo}
\end{equation}
where $E_{c}$ is the energy of the
cyclotron line.
The measured mass of the neutron star combined with theory for the internal structure
(\cite{baym79}) constrains the range for its radius.
If one assumes the magnetic field geometry at the
surface to be dipolar, then the size of the polar cap onto which matter
accretes is not well constrained by direct measurement.

%\placetable{tblcenx3}
EDITOR: PLACE TABLE 1 HERE.

Motivated by  earlier predictions of the likelihood of the
existence of Photon Bubble Oscillations (PBO) in Cen X-3
(\cite{klein96a}), we have analyzed two consecutive binary cycles
of the source to search for PBO. We have also calculated new
theoretical models constrained by the observable parameters of Cen X-3
%(Table~\ref{tblcenx3}).
(Table 1).
The value for $L_{x}$ is consistent with the observations reported here.
The physics describing the radiation
hydrodynamics which governs the accretion of matter onto the
highly magnetized polar caps of luminous X-ray pulsars, and which
has been incorporated in these new numerical simulations, has been
described elsewhere (\cite{arons87}; \cite{klein89}). Early
numerical results (\cite{klein91}) and linear stability analysis
(\cite{arons92}) suggested the formation of small scale but large
amplitude fluctuations in the matter density and velocity, which
form radiation filled pockets almost devoid of plasma
(``photon bubbles''). These photon bubbles (PB) are embedded in the
optically thick, inflowing plasma and as they grow non-linearly,
they result in significant observable fluctuations in the emitted
luminosity.

We observed the X-ray emission from Cen X-3 with the Rossi X-ray
Timing Explorer (RXTE) over two binary cycles of the source and
clearly detect both kHz QPO and a broadband kHz continuum in the
power density spectrum
which we identify with PBO and the high frequency power law continuum
generated by photon bubble ``turbulence''. (Our use of the
descriptive noun turbulence should not be confused with the
classical turbulence of a fluid.) Given
the observationally derived physical parameters for Cen X-3
which are summarized in
%Table~\ref{tblcenx3},
Table 1,
the {\it only} remaining
free parameter in our numerical models is the size of the polar
cap. In this paper, we show that by adopting a reasonable size for
the polar cap of Cen X-3, we are able to semi-quantitatively match
the frequencies of two observed QPO/PBO peaks
and the kHz continuum with the power density spectrum
of our calculated radiation-hydrodynamic models;
thus giving strong evidence for the existence of PBO and photon bubble
turbulence in Cen X-3.

\section{Observations}

Separate continuous observations of Cen X-3 during 1997
February-March (JD2450507.5067-JD2450509.2248;
JD2450509.5470-JD2450511.3304) were carried out with the PCA
on-board RXTE (\cite{bradt93} and \cite{swank97})
for two consecutive binary cycles of the source.
Each observation lasted 1.7 days; during the non-eclipse phases of
the binary orbit, data were acquired in two broad-band energy
channels at a time resolution of 62$~\mu$s. The observation modes
were specially selected to provide detailed information on the
high time resolution behavior of the 4.8 s pulse as well as any
broadband temporal behavior at frequencies above $\sim 100~$Hz.
Figure~\ref{figdata} is a log--log presentation of the average
deadtime corrected power density spectrum
for the entire data train of the low energy
channel (1--7 keV). The curve was computed by co-adding $\sim3400$
separate power density spectra, each with a $\sim65$ s exposure
and each composed of $2^{20}$ bins of duration 62 $\mu$s. These
power density spectra were further averaged over frequency bins
evenly spaced in logarithmic steps. This approach is the most
sensitive way to search for broadband features in the power
density spectrum over a wide range of frequencies
($10^{-2}-10^4~$Hz). Note the smooth plateau at very low
frequencies ($10^{-2}-10^{-1}~$Hz) followed by the first and
second harmonics of the 4.8 s pulse ($10^{-1}-10^0~$Hz). The
significance of the pulse is degraded by the crude averaging over
frequency. However, the integral of the power density still
correctly indicates the large pulsed fraction of the source. The
periodic pulse and its harmonics are superimposed on a broadband
power law which falls from an amplitude of $\sim10^{-1}$ to $10^{-6}$
with a slope of approximately $-2$ (see dashed line). At about
$100~$Hz another power density plateau appears, followed by two
QPO peaks ($\sim331$ and $\sim761~$Hz).
These two QPO peaks (50\% confidence ranges $260-407~$Hz and $671-849~$Hz)
are superimposed on the power density plateau which falls
sharply at frequencies above $\sim1200~$Hz. The curve falls to a
level of about $10^{-7}\left({\rm rms \over mean}\right)^2~{\rm
Hz}^{-1}$ at $\sim1800~$Hz which has reached the Poisson detection
limit of the transform. The high frequency Poisson limit in the form
a frequency dependent function has been
subtracted from the transform of the data (equation 6 of Appendix A).
The Leahy normalized version of the transform (\cite{leahy83}) was
carefully corrected for the effects of deadtime of the PCA (see Appendix A).

The same analysis procedure was applied to the high energy channel ($>7~$keV).
For these data the power density spectrum is nearly identical to that shown
in Figure~\ref{figdata} out to a frequency of about 50 Hz.
We detect no significant kHz features.
%The lower counting rate
%in the high energy channel precludes any sensitive measurement of the
%power density for frequencies above 100 Hz.
Determination of the energy dependence
of the kHz features of the power density spectrum must await significant additional
PCA data or future detections with instruments of much larger aperture than RXTE.

\placefigure{figdata}

We note that the identification of the specific features in the power density
spectrum as QPO in the presence of a significant continuum is a somewhat
arbitrary description. Others may prefer to identify the QPO as
``bumps'' in the continuum or just as a complex continuum. In this paper,
we choose to use the QPO designation in analogy with its use for the
discovery of QPO in the LMXB source GX5-1 (\cite{klis85a},\cite{klis85b}).
We justify this view based on the similar amplitude and width
of these QPO in the HMXB Cen X-3 and the low frequency ($\sim6~$Hz) QPO
in the LMXB GX5-1. Previous work on QPO in X-ray sources has indicated
a large dynamic range of Q values including examples of order unity.
The formal significance of the detection of these kHz QPO in Cen X-3
is not as high as for the QPO in GX5-1 but this fact is likely just a
coincidental result of the limited aperture of the PCA on-board RXTE.
Even if these kHz features are real statistical fluctuations in behavior
of Cen X-3, the QPO designation is still an appropriate description of the
results presented in this paper.
In using the QPO designation by observational analogy with GX 5-1
we do not suggest any underlying physical analogy.

A model composed of fifteen free parameters was fitted to the power
density over the frequency range $40-4000~$Hz. These parameters are
a low corner frequency $F_{l}$,
a high corner frequency $F_{h}$,
a level of power density $P_{l}$ at the frequency $F_{l}$
and its constant extension to lower frequencies,
the slope $S_{l}$ of a power law below the frequency
$F_{l}$ and its extension to higher frequencies,
the slope $S_{m}$ of a power law between the frequencies $F_{l}$ and $F_{h}$,
the slope $S_{h}$ of a power law above
the frequency $F_{h}$ and its extension to higher frequencies,
the frequency of the first Gaussian QPO
$F_{1}$, its rms amplitude $A_{1}$ and its width $W_{1}$, and the
frequency of the second Gaussian QPO $F_{2}$, its rms amplitude
$A_{2}$ and its width $W_{2}$.
Finally the last three parameters (A, B and C) model the frequency
dependent deadtime correction to the power density spectrum
(see Appendix A for details).

Alternative forms of the continuum power density will
slightly change the values of the six parameters that describe the
two QPO peaks but will not change the basic interpretation of the
fit. This particular functional form for this model was selected
as a reasonable choice for a simple characterization of the full
power density spectrum between $40-4000~$Hz with particular
attention to the frequency range $\sim100-2000~$Hz.

%Table~\ref{tbldata}
Table 2
displays a list of the parameters and their
best fitted values, as determined by a minimization of $\chi^2$.
The errors in the measured power density points were determined by
the Poisson counting noise. These data are comprised of 39
independent points, resulting in a minimum $\chi^2$ per degree of
freedom of 1.38 for 24 degrees of freedom.
Formally this is not an adequate fit, however
inspection of the residuals indicates that the fit is within
expected limits for frequencies between $100-2000~$Hz
(see Figure~\ref{figleahyerr} in Appendix A).

All components of the model match the data well
except for small deviations at low frequencies
(see Figure~\ref{figfit} and Figure~\ref{figleahyerr}).
The minor deviations from a power
law at low frequencies ($<100~$Hz) are revealed by the extremely
small error bars. These deviations which are a small fraction of the
power density at low frequency are likely caused by a real physical
mechanism that is more complex than a simple power law. Since the primary
goal of this paper is to characterize the kHz behavior of Cen X-3 we
did not further complicate the fifteen parameter model.
% The minor deviations from a power law at high
% frequencies ($>1200~$Hz) are an indication that such a simple model,
% although qualitatively correct, does not explain the fluctuations
% which likely indicate the presence of weaker unresolved QPOs at
% these higher frequencies.
The high frequency data are of
insufficient signal to noise to realistically determine the exact
nature of this detailed structure in the falling power density.
The uncertainty in the interpretation of the decline in power
above $\sim1200~$Hz is quantified by the large acceptable range of the
parameter $S_h$.
The presence of the two QPO peaks and the broadband kHz
continuum in the power spectrum over the frequency range
$100-1800~$Hz is highly significant.
Settting the amplitudes of the QPO peaks to zero results in an increase
in $\chi^2$ of $\sim47$ and $\sim42$ for QPO 1 and QPO 2 respectively.
This implies that the first and second QPO peaks are
formally detected at greater than $99.9\%$ confidence.
The lower and upper values of the parameters in
%table~\ref{tbldata}
table 2
are 50$\%$ confidence limits for each of the
fifteen free parameters of the model.
(These are determined by the variation of
each parameter until $\chi^{2}$ increased by $\sim14.34$, see \cite{lampton76}.)

The form of the fitted model for the low frequency ($40-100~$Hz) power law
was selected with the assumption that whatever the physical mechanism that
forms this power law that behavior is likely to continue past 100 Hz.
Since the extrapolation of the low frequency power law above 100 Hz is
much lower than the observed power density in the kHz band
this suggests a transition near $100~$Hz in which a new physical mechanism becomes dominant.
Note that the continuation of the power law to very low frequency
($<3~$Hz) rolls over to a flat level as required by causality since this
extrapolation cannot produce fluctuations that exceed 100$\%$ rms. This
damping of the behavior at low frequency is still consistent with a single
physical mechanism. Possibilities for the physics that generates this low frequency
power law are discussed in a later section.

%\placetable{tbldata}
EDITOR: PLACE TABLE 2 HERE.

Figure~\ref{figfit} shows the deadtime corrected data with $\pm 1 \sigma$ error bars
(dotted lines) as well as the model  (solid line). The data have
been multiplied by frequency in order to more clearly indicate the
Poisson counting noise. This also means that the vertical axis has
units of (rms/mean)$^2$ which allows for the easy approximate
interpretation of features in units of rms percentage amplitude.
Dashed horizontal lines at 2, 5 and 100\% are shown to indicate
the relative amplitudes of features of the power density spectrum.
The coherent pulsations from Cen X-3 clearly have a large
amplitude of order $\sim50\%$ whereas the high frequency
components ($>100~$Hz) have amplitudes in the 2-5$\%$ range. The
best fitted fifteen parameter model is also plotted in
Figure~\ref{figfit} and clearly tracks the measured data. Note the
presence of the two kHz QPO peaks ($\sim330~$Hz and $\sim760~$Hz)
and the break in the continuum
at the high corner frequency ($F_{h}\sim1200~$Hz).

\placefigure{figfit}

\section{Theoretical Model}

Photon bubbles instabilities can form in magnetized radiation pressure
supported atmospheres and subsonic flows (\cite{klein89},
\cite{arons92}) with $B$ fields as small as $10^8~$Gauss.
Classical X--ray pulsars such as Cen X-3 easily fulfill this
criterion.
We calculate a model appropriate for the surface conditions of the
neutron star Cen X-3. This model relies on the self-consistent
solution of the full two-dimensional, time-dependent equations of
the radiation hydrodynamics governing the accretion of matter onto
the highly magnetized polar caps of a luminous x-ray pulsar
(\cite{arons87}, \cite{klein89}, \cite{klein96a}). The full
description of the physics of highly magnetized flow that we have
included in our calculations is described in detail in
\cite{arons87}. We solve the full set of two-dimensional,
time-dependent relativistic radiation hydrodynamic equations.
These include the equations which govern the conservation of mass
and momentum, and separate energy equations for both the electrons
and the ions as well as a full non-LTE description of the time
dependent transport of radiation and photon number density
(\cite{klein91}, and \cite{klein96c}). We also include the full
effects of super strong magnetic fields on all physical processes
and we have developed a new flux-limited diffusion theory for the
transfer of radiation in optically thin, radiation filled
structures embedded in optically thick flows (Photon Bubbles).
We solve the entire
structure of self-consistent 2D coupled non-linear P.D.E's with a
2nd-order accurate, time explicit operator splitting approach.

The full evolution of the radiation hydrodynamic equations is an
initial value problem in time. The surface value of the magnetic
field strength assumed to have dipole geometry, the mass accretion
rate, and the size of the polar cap are all parameters, as are the
mass and radius (which enter into the strength of gravity) of the
neutron star.  We hold these characteristics fixed in the course
of a calculation (see Table 1 for values).
This leaves the
geometry of the polar cap as the {\em only} free parameter in
the calculation.  In principle, a specification of the surface
field strength $B_0$ and the intrinsic luminosity $L_x$ could,
under certain assumptions (i.e., magnetic dipole field
configuration) determine the radius of the polar cap $\theta_{c}$
(\cite{pringle72}, \cite{davidson73b}, \cite{lamb73},
\cite{arons80}).
The details of the magnetospheric mass loading are quite uncertain.
Since these details determine the size of the polar cap and the
distribution of
the accretion mass flux within the polar flux tube, we simplify by assuming
the accretion flux to be constant from the magnetic axis out to a fixed
magnetic flux surface, then zero beyond, and we take the polar cap radius
to be
the only free parameter in our calculations.

From our previous calculations with $L_x = 3\times 10^{37}$ ergs
s$^{-1}$ and $B=3\times 10^{12}~$G, we know that in models with
large polar caps ($\theta_{c}=0.4$) we obtain two PBOs with
frequencies of 170 Hz and 260 Hz (\cite{klein96b}) while smaller
polar caps ($\theta_c=0.1$) yield PBO with significantly higher
frequencies in the range of 3500 Hz (\cite {klein96a}). Thus wider
accretion columns produce PBO of lower frequency, a scaling most
likely related to the longer diffusion time necessary for photons
which originate in interior photon bubbles of wider columns to
diffuse to the outer edge of the accretion column. Therefore, PBO
at the observed frequencies at 330 Hz and 760 Hz should occur for
polar cap sizes bounded by our previous calculations, $0.1 <
\theta_c < 0.4$. We have computed three new models to explore the
variation of PBO frequency with polar cap sizes $\theta_c$ = 0.25,
0.3 and 0.4, with the luminosity $L_x =  4.7\times 10^{37}$ ergs
s$^{-1}$ and magnetic field strength $B = 3.2\times 10^{12}$ G
inferred from observations of Cen X-3. The value of $L_x$ selected
here is one half of that shown in
%table~\ref{tblcenx3}
Table 1
which is appropriate for one of the two polar caps.
The calculations refer to accretion onto a single polar cap.
%Clearly, we would like to compute models which fully sample polar cap
%sizes $\theta_c$ from 0.05 to 0.5 radians. The restriction of reporting
%on only three new models in addition to those presented in previous
%publications (\cite{klein96a} and \cite{klein96b}) is due to
%limited time and finite computational resources.
%A full exploration of the
%parameter $\theta_c$ and variations in the other physical parameters
%is underway but will require 1-2 years of research.

Since the basic oscillation and instability time scales of the
magnetically modified g-modes which become photon bubbles are
proportional to the time scale for radiation diffusion
(\cite{arons92}), and photon escape from the {\it sides} of the
column controls the  emergent intensity, one can reasonably
suppose that the time scale for radiation diffusion across the
magnetic field, $t_{diff} \simeq L_\perp^2 (3/c \lambda_{mfp,
\perp}) = L_\perp /V_{diff\perp}
\propto \theta_c^2$,
might be related to PBO timescales. This strong dependence
on $\theta_c$ is amply supported by the simulations previously published and
reported here. Here $V_{diff\perp} = c /3 \tau_\perp$
 is the diffusion velocity across
the flow above the accretion shock, with $\tau_\perp$ the
scattering optical depth across the magnetic field. $L_\perp$ is
the transverse distance across the magnetic field
 which
at the base of the accretion mound on the stellar surface
is $R_\star \theta_c$, but further up in the accretion column
is $R_\star \theta_c (r/R_\star)^{1.5}$.
The parameter $\lambda_{mfp,\perp}$ is the transverse mean free path across the
accretion mound.
We find in the $\theta_c = 0.25$ model that just above the
accretion shock at the accretion mound, $\tau_\perp$ = 19 and
$L_\perp$ = $6\times 10^5~$cm, yielding $t_{diff}$ = $1.14 \times
10^{-3}~$s.  This corresponds to a diffusion frequency $\nu_{diff}
= 877~$Hz.  At the base of the accretion column, we find
$\tau_\perp$ = 80 yielding $t_{diff}$ = $2 \times 10^{-3}$ s and
the corresponding diffusion frequency $\nu_{diff} = 500~$Hz.

Calculating the transverse radiation transport speed as
$F_\perp/E_{rad}$ at the accretion shock, where $F_\perp$ is the
transverse radiation flux and $E_{rad}$ is the radiation energy
density in the $\theta_c = 0.25$ model, and setting the result
equal to $V_{diff}$, gives a diffusion speed of $1.0\times 10^9$
cm s$^{-1}$ and a  diffusion time $t_{diff} = 6.04\times 10^{-4}$
s. This corresponds to a diffusion frequency $\nu_{diff} =
1.6\times 10^3$ Hz.  Taking the measurement of the transverse
radiation transport speed near the base of the accretion column
gives $V_{diff} = 5\times 10^8$ cm s$^{-1}$, $t_{diff} = 5\times
10^{-4}$ s and $\nu_{diff} = 2000$ Hz.  Thus the two ways of
computing the diffusion time are in agreement within roughly a
factor of 2.
%A Monte Carlo treatment following the time history
%of the photons in the accretion column would be necessary to get a
%precise determination of the diffusion time.
Considering these
approximations, one might expect this model to show PBO with
fundamental frequency $\nu_{PBO} \sim 1000$ Hz, not drastically
different from the values found in the computational model and
in the observations of Cen X-3. 
The uncertainty in the quantification of the diffusion time to within
a factor of 2 is a consequence of the highly non-uniform structure of the
accretion column.  If the column were uniformly dense, the two approximations
for $\tau_{diff}$ would be in more precise agreement.  This factor of 2
uncertainty has no bearing on the accuracy of the PBO centroid
frequencies which are determined by the solutions of the full
radiation hydrodynamics equations in a spatially inhomogenious medium.
Thus the final PBO centroid frequencies represent a complex
path for photons through intermingled optically thick and
thin plasma.  A diffusion time estimated at a few points in such a medium
is not likely to be better than a factor of 2 in accuracy.  A Monte
Carlo treatment following the time history of the photons
in the accretion column would be necessary to get a precise determination
of the diffusion time for photons to leak out of the column.
%The uncertainty in the quantification
%of the diffusion time is a consequence of the inadequacy
%of any summary of a complex system by a single average
%parameter. This limitation of physical interpretation should not
%be confused with any lack of robustness of the full numerical models.
%Independently, future research is necessary to fully understand the
%quantitative limitations of the numerical models including a precise
%evaluation of the sensitivity of the frequencies of the PBO to
%changes in the parameters of the physical model or parameters of the
%numerical methods.

We identify the high corner frequency $\sim 1200~$Hz noted
previously with an average diffusion time scale $t_{diff}$
which we relate to the total optical depth $\tau_{obs}$
across the accretion mound.
\begin{equation}
   \tau_{obs}~=~26~
   {\left[ {\left( F_{h} \over 1200~{\rm Hz} \right) }
           {\left( {{(1 + Z)} \over {1.3}} \right) } \right]}^{-1}
   {\left[ \theta_{c} \over 0.25~{\rm radians} \right] }^{-1}
   {\left[ R_\star \over 10~{\rm km} \right] }^{-1}
\label{diffusion}
\end{equation}
This equation for $\tau_{obs}$ is a recasting of the expression
for the diffusion time $t_{diff}$, now expressed in terms of the
optical depth. The form of Equation~\ref{diffusion} indicates how
the value of $\tau_{obs}$ varies depending on the assumed size of
the polar cap, $\theta_{c}$. The particular value of $\theta_c$ is the
assumed value that we used for model A which also corresponds to a
typical value within the expected range for $\theta_{c}$ (see
%Table~\ref{tblcenx3}).
Table 1).
Basically any internal oscillations within
the interior of the accretion mound which are faster than the high
corner frequency are suppressed by a factor that scales
approximately as the inverse second power of frequency. This
simple scaling is the result of the approximation that diffusion
is just a smoothing process by a linear filter with a
characteristic time scale. The estimated value of the optical
depth $\tau_{obs} \sim 26$ agrees approximately with the values of
optical depth in the numerical calculations (typically $\sim20-80$).
This agreement is not
expected to be better than a factor of a few because of the
simplifying assumption of a uniform medium with a diffusion
velocity $\left( c \over 3 \tau_{obs} \right)$.
We identify this feature at high frequency as the first direct evidence
of radiative diffusion near the surface of a neutron star (NS).

\section{Results of Radiation Hydrodynamic Calculations}

Figure~\ref{figmodel} shows the results for the power density
spectra of the time series of the emergent transverse luminosity
out of the sides of the accretion column for three models (A, B
and C) for which the values $L_{x}$ and $B$ have been set to those
measured for Cen X-3 (see
%Table~\ref{tblcenx3}).
Table 1).
The models A, B
and C are distinguished by setting the polar cap size to 0.25, 0.3
and 0.4 radians respectively.
The power density
spectrum for each model is obtained by first performing a full 2-D
time-dependent radiation-hydrodynamic calculation and obtaining
the radiative flux in the transverse direction.  This flux is
integrated along a series of annular rings along the sides of the
accretion funnel to obtain the emergent luminosity which exits the
accretion column in a low altitude fan beam of radiation.  The
time series of the transverse luminosity is then Fourier
transformed to obtain the power density spectrum.

\placefigure{figmodel}

These models are simulations of only a few tens of milliseconds of
accretion onto the neutron star and may not be well defined ergodic
samples necessary for an exact comparison with the observed power
density spectra averaged over many thousands of seconds. However,
in support of making such comparisons of simulations to
observations, we point out that any initial transients in the
calculations die out after less than $10\%$ of the running time for each
simulation. The power spectra calculated from the simulations are
derived from the remaining $\sim 90 \%$ of each run, corresponding
to what appears to be a statistically steady state. Our procedure
limits the quality factor, Q, of any individual PBO to the ratio
of the simulation's length to a given PBO period ($Q  < 100$). Our
procedure obviously would miss slow changes of the apparent
statistically steady states observed. We have no reason to believe
that there are such slow systematic effects, given the physics
incorporated in the simulations. Therefore, we take the
simulations as our best known estimates of the behavior of PBO
over times much longer than characteristic PBO periods. We further
note that the errors on the actual calculation of the power
densities are very small (less than $10^{-8}$) in the units shown
in figure~\ref{figmodel}. The detailed features in the power
densities are formally {\em very} significant with none of the
degradation at high frequencies that typify observations
containing Poisson counting noise. These very small errors, which
cannot practically be shown in Figure~\ref{figmodel}, are due to
the finite spatial grid and discrete time steps of the numerical
calculations.
This level of this numerical noise in the power density
is revealed by an artificial downturn at
very high frequencies ($\sim10^{5}~$Hz).
Typically, timesteps in the calculations are $\leq
5\times 10^{-5}/\nu_{PBO}$.

In Figure~\ref{figmodel}, the lower curve is model A ($\theta_c =
0.25$ radians), the middle curve is model B ($\theta_c = 0.3$
radians) and the upper curve is model C ($\theta_c = 0.4$
radians). Models B and C have been artificially shifted by two and
four decades vertically for illustrative purposes. The curves in
the left panel are shown in high resolution to reveal the details
of structure of the power density spectrum.
Below a frequency of $\sim1~$kHz the spectrum is shown at the
full resolution of the calculations. At higher frequencies the
data are smoothed in equal logarithmic intervals for clarity of
presentation.
The curves in the
right panel show the same data further smoothed to facilitate comparison
with observations (see figure~\ref{figfit})
and to enhance the broadband features. The most
striking common feature of all three models is the rolloff at high
frequencies in approximate agreement with
Equation \ref{diffusion}
and a clear demonstration of the signature of radiative diffusion.
In detail Models A and B show a somewhat steeper slope at higher frequencies
which is not are that unlikely considering the rough approximation
of Equation \ref{diffusion}.
Each curve has a principal low frequency PBO that is shifted to
higher frequency for the models with smaller polar caps in good
agreement with $\nu_{PBO} \propto 1/\theta_c^2$ scaling. The lower
frequency PBO in model A determined from the detailed calculations
is $\sim350~$Hz (not redshifted), which agrees well with the observed lower
frequency QPO for Cen X-3 ($\sim~$330 Hz).  Our detailed
calculations yield lower frequency PBO approximately  at 250 Hz
and 200 Hz for models B and C with $\theta_c = 0.30$ radians and
$\theta_c = 0.40$ radians respectively.  These are in reasonable
agreement with the frequencies  246 Hz and 138 Hz suggested by
$\theta_c^2$ scaling. Also the level of the continuum is somewhat
lower for models with smaller polar caps.

Note that the absolute levels of the power density can be directly
compared to the observed levels in Cen X-3 (see
Figure~\ref{figfit}). The amplitude of the observed power density
in the range $100-1800$ Hz is about one or two orders of magnitude
higher than the level calculated in the theoretical models. This
discrepancy is not that severe considering the simplified model
for the geometry of the emission region.
%and that the full range of
%observed power density ranges over 7 orders of
%magnitude (see Figure~\ref{figdata}).
The actual rms amplitudes of
the fluctuations in the models near 1 kHz are only a factor of
ten lower than the observed level. The variability of the escaping X-ray
luminosity is a damped form of the more violent internal photon
bubble variability deep within the accretion mound. The presence
of the PBs within the mound necessarily implies the local internal
opacity variations have rms amplitudes that reach nearly $100\%$.
In an extension of the 2D theoretical models to 3D we would expect
the variations of opacities ``seen'' by internally generated photons to
vary by an even wider range. Some of these photons might ``see''
reduced opacities episodically which would significantly enhance
the X-ray variability and rms amplitudes seen by a distant
observer. Also the assumption of a uniformly filled accretion column
might be relaxed  as a partially filled column (hollow cone) would also
provide lower opacity paths from the interior of the shock mound
to the exterior surroundings.  We are extending our calculations to a
hollow cone accretion 
geometry as the first step toward investigating 3D effects
on the amplitude of the PBO fluctuations.  This work will be reported
elsewhere.   There are likely very low frequency
variations ($<~1~$s or pulsed modulated at 4.8 s) in the geometry
of the matter filling the accretion column that would result in low
opacity ``windows'' which would allow a distant observer to ``see'' into
the interior of the shock mound where the rms fluctuations of the
X-ray emission are much higher.

Model A ($\theta_c = 0.25$ radians) shows PBO at 350 Hz and a
composite PBO at about 700 Hz.  The frequencies of these PBO are
in good agreement with frequencies of the QPO peaks observed in the
power density spectrum of Cen X-3.
The relative amplitudes of the PBO/QPO as compared to the
continuum power are a better match between the theory and
observation than the absolute amplitudes.
The rms amplitudes of the observed QPO 1 and 2 in
%table~\ref{tbldata}
Table 2
can be expressed as
equivalent widths ($\sim710~$Hz and $\sim970~$Hz) which are comparable
to the full range of the continuum ($\sim100~$Hz to $\sim1800~$Hz)
indicating an approximately equal contribution of the QPOs and the
continuum to the full rms amplitude above $\sim100~$Hz.
The approximate balance between PBO/QPOs and continuum is also roughly true
for all three theoretical models.
The additional strong
broad composite PBO at $\sim3000~$Hz that is seen in models A and B is not
seen in the observed power density spectrum of Cen X-3. However it
is below the level that could be easily detected with the PCA. The
upper limit for the detection of broad features in the power
density spectrum rises proportional to frequency above
$\sim1000~$Hz which rapidly decreases the capability for detection
of any features near $3000~$Hz. This rapidly declining ability to
detect broad features at high frequencies should not be confused
with the flat asymptotic behavior for the detection of features of
fixed bandwidth that have increasing Q values at higher frequencies.

Models A, B and C have been plotted
simultaneously with the observed data in Figure~\ref{figcompare}.
For comparison both models and the observed data have been plotted
over an appropriate frequency range and on the same absolute scale
with no effort to adjust the amplitudes to match the data.
The frequencies of the theoretical simulations have been red shifted correctly
by the required factor of $(1 + Z)=1.3$ appropriate for the
surface of a neutron star with mass and radius specified in Table 1.
The frequencies of the PBO in model A match the frequencies of the
QPO in the observed data.
This is remarkable considering that
the only free parameter in our calculations is the size of the polar cap.
The theoretically determined PBO
at $\sim300~$Hz and the corresponding observed QPO at a
similar frequency are both relatively weak in comparison to the
PBO/QPO at $\sim700~$Hz. In addition model A (also B) has a broad PBO
with a large amplitude at $\sim3000~$Hz that is very near the
limit of the observable range of RXTE (Figure~\ref{figmodel}).
The continuum shape of model A does not match the observations as
well as Model C (Figure~\ref{figcompare}).
All three models show a power law rolloff at high frequency ($\sim10^4~$Hz)
in agreement with the simple approximation that diffusion is
just a smoothing process by a linear filter with behavior $f^{-2}$.
Models A and B show a somewhat faster rolloff at even higher frequencies.
If we consider the whole picture, model A has the best QPO/PBO frequency
match and models B or C have a better match to the continuum.
These models taken together determine a range for the polar cap
radius of $\sim$2--3 km.  Calculations are underway (to be reported elsewhere)
to better bracket the polar cap radius  by
doing a parameter study of small variation of the polar cap radius around
our best model A with $\theta_c = 0.25$ radians, and its effect on the PBO frequencies.  
These calculations are extremely time 
consuming and typically require $\sim$ 300 hours per model on a 
supercomputer.  

\placefigure{figcompare}

Our calculations have used measured parameters for Cen X-3 (e.g.,
luminosity, magnetic field, mass, etc.) while only varying the
size of the polar cap.  Hence the identification of the kHz QPOs
and a kHz broadband continuum with PB phenomena suggests
that the radius of the polar cap of Cen X-3 is
``measured'' as $\sim$2--3 km. This is the first, albeit model
dependent, measurement of the size of the accreting polar cap of
an X-ray pulsar.

\section{Discussion of Low Frequency Behavior ($<100~$Hz)}

A simple power law with a slope $\sim-2$ matches the
broadband continuum from $3-100~$Hz in the power density spectrum of Cen X-3
(see Figure~\ref{figdata}, Figure~\ref{figfit} and
%table~\ref{tbldata}).
Table 2).
This general behavior has been previously reported for Cen X-3
and other HMXB sources (\cite{belloni90}).

The origin of this lower frequency noise may be the consequence of the physics 
controlling the loading of mass onto the magnetospheric field, perhaps
modulated by intrinsic turbulence in the disk. In the case of Cen X-3, the mass
flow exterior to the magnetosphere is almost certainly in an accretion disk.
To fix ideas, suppose the
boundary layer between the disk and the magnetosphere is Rayleigh-Taylor
unstable, a mass entry mechanism previously treated in some detail for the case 
of spherical accretion by \cite{arons76}, \cite{arons80}, \cite{elsner77},
\cite{elsner84}.  Some aspects of Rayleigh-Taylor instability in the
disk/magnetosphere interaction have also been discussed by \cite{taam90}.
 
The thickness and height $H = R_m (c_{ms} /v_K)$ of the disk's boundary layer 
set the size of an unstable mode (a ``blob'').  Then
$\delta M \sim \rho H^3$ is the mass of a blob, while in statistically steady
accretion $\rho = \dot{M}/2\pi H R_m v_K$ (see \cite{arons93} for this result,
in the relevant case when the plasma is diamagnetic, and \cite{ghosh79}
when macroscopic dissipation is assumed to give the free access of the
neutron star's magnetic field to the disk's plasma on microscopic scales.) Here 
$R_m \approx 4300$ km is the magnetopause radius, evaluated assuming
approximate corotation between the Keplerian disk's inner edge and the 
magnetosphere (\cite{bildsten97}), 
$v_K \approx 5580$ km/s is the Kepler velocity of the disk just outside the 
magnetopause, and $c_{ms} = (c_s^2 + v_A^2)^{1/2}$ is the magnetosonic speed
in the disk, with $c_s$ the sound speed and $v_A$ the Alfven speed determined
by the disk's own magnetic field.  
Compton scattering of the hard X-rays from the star limits the sound
speed to $c_s \sim 1000~{\rm km}~{\rm s^{-1}}$.  As we shall show, in mass transfer by blobs,
the disk's inner edge must have a strong magnetic field of its own, such that
$v_A^2 \gg c_s^2$. 

The mass of a blob is $\delta M = \rho H^3 \sim \dot{M} H^2/2\pi R_m v_K$,
and the number of blobs contributing to the mass flow is
$N_{blob} \approx 2 \pi R_m /H$.  Therefore, in a statistically steady
state,
\begin{equation}
\dot{M} = \frac{d}{dt} N_{blob} \delta{M} \sim 
  \frac{N_{blob} \delta{M}}{\tau t_{ff}} = \frac{\sqrt{2}}{\tau} \dot{M} \frac{H}{R_m}.
 \label{eq:blobtransf}
\end{equation}
Here $\tau$ is the number of magnetospheric free fall times 
$t_{ff} = R_m /v_{ff} (R_m) = (R_m^3 /2GM_*)^{1/2} \sim 0.46~$s
required for a blob to form, decelerate and freely fall through a
distance $\sim H$, thus
detaching from the boundary layer. This detachment process is very ill
understood. If one assumes the blob loses its orbital velocity instantaneously,
then $\tau =4 \; (H/R_m \ll 1)$, while $\tau \approx 3$ if $H/R_m \sim 1$.
$\tau $ is expected to be larger still, if one accounts for the mechanical
and electromagnetic drag forces on the forming blob.  

Eliminating $\dot{M}$ from \ref{eq:blobtransf} yields 
%\frac{H}{R_m} = \frac{c_{ms}}{v_K} \approx \frac{\tau}{\sqrt{2}} \sim 2.
\begin{equation}
\frac{H}{R_m} = \frac{c_{ms}}{v_K} \approx \frac{\tau}{\sqrt{2}} \approx \frac{3}{\sqrt{2}} \sim 2.
\end{equation}
Therefore, the inner edge of the disk must be reasonably thick, with 
$N_{blob} \sim 2 \sqrt{2}\pi / \tau \sim 3$ blobs contributing at any one
time to the mass flow, and the magnetic pressure in the disk must be
large compared to the gas pressure. For Cen X-3's parameters,
we find $B^2 /4\pi \sim 230 nkT_{gas}$. Note that we have neglected the
inflationary effects of radiation pressure on the disk. Since Cen X-3's
luminosity is $\sim$ 30 \% of the Eddington luminosity, some of the
disk's inflation might be supported radiatively rather than magnetically,
but it is hard to avoid the conclusion that magnetic support makes a
large and probably dominant contribution to the disk's inner
structure.

Such variability in the accretion flow will cause the size of the polar cap
to vary. In the ``kinematic'' estimates of the size of the polar cap as a
function of $\dot{M}$ (\cite{pringle72}, \cite{davidson73b}, \cite{lamb73}), 
$\theta_c \propto (R_* /R_m)^{1/2} \propto \dot{M}^{1/7}$. Such a model might
apply if a) macroscopic reconnection causes all dipole
field lines which would have closed at radii $r > R_m$ to be open and
b) blobs do not fall to radii substantially smaller than $R_m$ before 
their particles become frozen to magnetospheric field lines. If, on
 the other hand, blobs form diamagnetically, lose their angular momentum
and fall freely before disrupting into particles which freeze to the
magnetospheric field, then $\theta_c \propto \dot{M}^{-2/7}$ (\cite{arons80}). 
 If blob formation causes the mass flux onto one pole to vary by as much
as $\sqrt{1/(N_{blob} /2)}$, then in Cen X-3, the polar cap size might
be varying by as much as 12 \%  (open field model) or 23 \%
(freely falling blob model) over the course of the three days
of observation that went into forming the power spectrum. In principle,
we should compare the average of several models similar to model A
but with different values of $\theta_c$ to the high frequency fluctuation
power, rather than the single ``snapshots'' shown in models A, B and C.
However, the expected small range of variation in $\theta_c$ suggests
the snapshot comparison to be good enough, given the
limitations in the simulations.

Consideration of Kelvin Helmholtz mixing ({\it e.g.} \cite{burnard83});
lead to similar 
conclusions. The KH effects most likely contribute to blob formation and 
deceleration, while Rayleigh-Taylor (RT) instabilities
control the detachment of blobs
from the disk's inner edge.  For both effects, the basic mode size is
$H$, and the mass transfer rate is controlled by gravitational free
fall, so our order of magnitude estimates are invariant to the greater
complexity of considering both RT and KH together. On the other hand,
if the mass transfer were to be controlled solely by reconnection,
with a mass loading velocity $\sim \epsilon v_A$ with $\epsilon$ perhaps
on the order of 0.1 (see \cite{lazarian99} for a recent investigation
of rapid reconnection and full references on the subject). Repetition
of our arguments lead to $H/R_m \sim 2\pi /\epsilon \gg 1$, which
is impossible since
the magnetic stresses would cause the disk to explode and shut off accretion.

These estimates lead to several important conclusions. Only 
$N_{blob} \sim 3$ ``blobs'' contribute to the mass flow onto each polar cap
at one time, with the basic time scale for these slow fluctuations
being $\tau t_{ff} \sim 1.4$ seconds.  Thus, the observed fact that
spectrum flattens at less than 3 Hz can be understood as the consequence of the
fluctuations becoming large amplitude at these low frequencies.
Possibly 
additional fluctuations are imposed on the structure of the blobs by
intrinsic disk turbulence.
Hawley has calculated models of disk turbulence which show a
scale free power law dependence of the spatial transform of density,
velocity and kinetic energy of the flow (\cite{hawley95}).
The necessary truncation of the
power law on the largest scale corresponds to overturning velocity structure
of order the inner disk scale height. These slowest components of the flow
could naturally explain the flattening of the temporal transform
at low frequency ($<3~$Hz).
The power law form of the spatial transform would naturally imply a power
law form of the temporal transform.
The reality of these fluctuations
depends on the assumption that the disk
variability imposes a structure on the formation of blobs.
Further, since
the Rayleigh Taylor modes can include structure elongated
along the circumference of the magnetosphere (``blob'' azimuthal lengths between 
$\sim H$ and $\sim 2\pi R_m$),
lower frequency power might also be caused by modulations of the flow due to 
azimauthal structure in the instabilities, especially if one takes account
of the time needed to get rid of the orbital velocity and allow a blob to go 
into free fall. Quantitative consideration of these more advanced
issues is far beyond the scope of this paper. However, whatever mechanism modulates
the accretion rate and accounts for the low frequency ($3-100~$Hz) power law with
a slope $\sim-2$, its behavior is likely to extrapolate as a continuation of
the same power law to higher frequency ($>100~$Hz). Therefore the deviation from this power
law at $>100~$Hz that is observed in Cen X-3
is likely a different physical mechanism associated with the surface of the neutron star.

\section{Conclusions}

We have advanced the view that the fluctuations observed in the
power spectrum at frequencies exceeding a few hundred Hz have their
origin in the intrinsic photon bubble instability of the accretion flow
at the stellar surface, rather than being the passive result of still higher
frequency fluctuations in the mass loading into the magnetsophere. Emprically,
this view is supported by the change in the power spectrum's structure above
a few hundred Hz. From the point of view of polar cap plasma dynamics,
the large amplitude fluctuations in the mass flux implied by the unstable
mass entry model outlined in the previous section
can be modeled as a steady flow, since the
time variation scale of the mass flux ($\sim 0.5$ sec for Cen X-3) is far
longer than the intrinsic photon bubble instability time scale (milliseconds).
Therefore, our radiation hydrodynamic calculations, which assume steady
flow on the photon bubble time scale, are directly applicable. Our
identification of the high frequency fluctuations in Cen X-3
(kHz QPO and broadband kHz continuum) with the
photon bubbles (PBO and turbulence) is well supported by the fact that the
four parameter radiation hydrodynamical simulations match the
observations remarkably well in the applicable frequency range ($>100~$Hz),
with an adjustable polar cap size $\theta_c$
as the only free parameter of the fit.
The mass of the neutron star, the
magnetic field at its surface and the accretion rate are all
constrained by other observations. 
Adjusting $\theta_c$ yields simulation calculations that qualitatively and
approximately quantitatively explain eleven (of fifteen) free parameters of
the best fit to the observed data. These include the form of the
continuum power with a plateau, the high corner frequency and
rolloff power law and the presence of two broad lines (QPO) at the
appropriate frequencies, amplitudes and widths. The remaining four
parameters of the fit to the observed data concern the low
frequency ($<100~$Hz) behavior of the source
and the PCA deadtime corrections.
The range of applicability of the calculations
does not include behavior of the source for frequencies below $\sim100~$Hz.
The mass entry model
outlined above suggests that these low frequency effects are related to
modulation of the accretion rate due to the intrinsic instability
of the mass entry mechanism, to which the X-ray emission from the surface
passively responds.  Thus, the match of the models to the observations 
lends strong support that our discovery of kHz fluctuations in the
X-rays emitted by Cen X-3 can be identified with our prior suggestion
that PBO would be observed in this X-ray source (\cite{klein96a}).

The similarity between GRO J1744-28 and Cen X-3 is significant and
strengthens the PBO identification for GRO J1744-28
(\cite{klein96b}). Both GRO J1744-28 and Cen X-3 show two QPO
peaks and a strong continuum in their power density spectra
(\cite{zhang96}). The less complex pulse profile of GRO J1744-28
(nearly a perfect sinusoid) suggests a larger polar cap than for
Cen X-3 in agreement with observations of lower frequency PBO/QPO.
It is clear that the $\sim400~$Hz separation in the kHz QPOs in
Cen X-3 has nothing to do with rotation since this separation
frequency is vastly different than the 0.21 Hz spin frequency.

In contrast with the interpretation of GRO J1744-28,
the identification of photon bubble phenomena in CenX-3 is much
firmer because of the naturally faster time scales extending to
$\sim1800~$Hz in the broad continuum. The observations reported here
reaching to $\sim1800~$Hz are the fastest known variations in the X-ray
emission from an accreting compact object. We can not exclude the
simultaneous presence of additional mechanisms for temporal
variability above $100~$Hz but PB models provide the only
mechanism which yields quantitative estimates for both amplitudes
and frequencies.

Although the PB simulations include most of the accretion physics 
thought to be important
near the surface of a neutron star undergoing optically thick accretion
at the surface, there are some practical limitations. These models can
currently only be integrated for a few tens of milliseconds, which
limits a detailed comparison with averaged power density spectra
over days of exposure in an observed source.
Future X-ray missions with significantly larger area detectors and
parallelization of our algorithms would circumvent these
difficulties.

Prior to the RXTE launch we predicted the observation of photon
bubble phenomena including the specific existence of kHz fluctuations and
some power in quasi-periodic oscillations identifiable
as PBO in Cen X-3 (\cite{klein96a}). We have now
observed these fluctuations in Cen X-3 and have refined the model
calculations to match the actual frequency dependence of the power
density spectra including the presence of a significant continuum.
A careful comparison of the predicted power density from the PB
simulations with the observed power density spectrum constrains the size of the
polar caps of Cen X-3 to a radius of $\sim$2--3 km. This  has
permitted the first model-dependent measurement of the size of the accreting polar
cap in an X-ray pulsar.

The identification of PB phenomena in Cen X-3 opens the door to
future X-ray observations with high resolution timing that will allow us to probe the
physics of the surface of super-Eddington accreting X-ray pulsars.
\section{Acknowledgments}

We acknowledge helpful conversations with Lars Bildsten, Lynn Cominsky,
Saul Rappaport and Will Zhang.
We thank Ed Morgan for his extensive independent appraisal of the data
analysis portion of the paper. We also thank both Ed Morgan and Wei Cui
for providing the GX 17+2 data and the Cyg X-1 data respectively.
These data were crucial for the development of a correct deadtime model
of the PCA.
This research has been supported by the award of a NASA ATP grant NAG5-3809 and
a NASA XTE Guest Observer grant NAG5-3385.
Part of this work was supported under
the auspices of the US Department of Energy at the
Lawrence Livermore National Laboratory
under contract W-7405-Eng-48.
The numerical models were calculated with the J90 supercomputers at LLNL.

\clearpage

\section{Appendix A: PCA Deadtime Corrections}

In this appendix we describe the procedure for correcting the power density
spectrum for deadtime effects in the PCA.
These corrections are required and are important in those cases in which the
Leahy normalized power density falls below the level of 2
that is expected if no deadtime effects are present.
We use a frequency dependent model for an additive deadtime component
of the power density of a form given by 
equation~\ref{deadtime}. This Leahy normalized deadtime model $P_{dt}(f)$
is suggested by equation (22) of \cite{zhang95}
and equation (2) of \cite{zhang96} where f is the frequency.
The functional form for the paralyzable and nonparalyzable deadtime models
are nearly the same (compare Figures 1 and 3 in \cite{zhang95}).
The practical distinction between the two models
is insignificant below $10~$kHz.
The effect of discrete time binning is also unimportant below $10~$kHz for
the data shown in this appendix.
The value of $t_d$ is about $10\mu$s and is a fixed property of the
function of each of the five modules that comprise the PCA.
The value of $t_{vle}$ is a selectable parameter of the
flight configuration of the PCA and is
$\sim61\mu s~$(level 1)
or $\sim142\mu s~$(level 2) for all data
shown in this appendix. These selectable values of $t_{vle}$ are averages of
separate ground calibration measurements for the five PCUs of the PCA. The variations
in these values of $t_{vle}$ are of order $1\%$ which is more that sufficient
for the purpose of evaluating deadtime corrections.
%Precise knowledge of the fixed values of $t_d$ and $t_{vle}$ is not required.
The variables $r_o$ and $r_{in}$ are the
actual counting rate and the incident counting rate (assuming no
deadtime correction) respectively.
The variable $r_{vle}$ is the total rate of VLE (very large event) due to
background particles and pileup of X-rays. 
These rates correspond to the values for one of the five PCA modules.
Clearly the values of $r_o$ and $r_{in}$ are complex time dependent
functions as revealed by the actual observations of $r_o$.
Equation~\ref{deadtime} is suggested by the simple situation in which
both $r_o$ and $r_{in}$ are constant.
With further simplifying assumptions that $r_o~\approx~r_{in}$
and $r_o~\ll~t_d^{-1}$, we can express the values
of A and B as approximate functions of the rates $r_o$ and $r_{vle}$
(see equations~\ref{Adead} and ~\ref{Bvle}).
We will show that the overall deadtime correction $P_{dt}(f)$ can be
successfully represented by fixed values of A and B for actual observations.
Note that our proposed procedure for estimating the deadtime corrections is
to vary the values of A and B as additional free parameters of any model
that one might use to fit a power density spectrum. The approximate relationship 
of A and B to any rates ($r_o$, $r_{in}$ or $r_{vle}$)
as given by equations~\ref{Adead} and \ref{Bvle} serves only as a verification for sensible
values of A and B since these equations are based on a model that assumes that the
rates are constant when in reality these rates are highly variable. 
Parameter C and third term of equation\ref{deadtime}
is the additive component of
background present in all PCA observations. Its functional form was determined by
approximately (within about $1\%$) fitting the empirical power density spectrum of
a deep exposure to background with no signicant sources present (\cite{morgan99}).
This frequency dependent function shows clear minima at 2000, 6000 and
10000 Hz and maxima at 0, 4000, and 8000 Hz. These features are likely produced by
background events in channels 0-7.
The total count rate from the background for GX 17+2, Cyg X-1 and Cen X-3 in channels 0-17 is
$\sim30~s^{-1}$.
Since the background rate
is not separately measured in the presence of a bright source there is no firm
estimate of the background rate during an typical observation. The conservative
procedure is to include an additive  term in the fit to any observe power spectra with an
amplitude proportional to the parameter C. This background term is not
really a deadtime correction but is encluded in the procedure since it is
a model for the frequency dependent form of a needed correction.
In the worst case an error in the determination of the values of A, B and C would correlate with some
real component of the power density spectrum of an observed source. This possibility always leads to
an underestimate of the actual power density spectrum of the source. Therefore this
procedure which freely varies the parameters A, B and C is a conservative approach for
estimating any components of the power density spectrum which are not real components of
the observed target.

\begin{equation}
   P_{dt} \left( f \right)~=~2~-~A~
     {\left[ sin \left( 2 \pi t_d f \right) \over
                  2 \pi t_d f \right]} 
     +~B~
     {\left[ sin \left( \pi t_{vle} f \right) \over
                      \pi t_{vle} f \right]}^{2}
     +~0.8744~C~
    {\left[ e^{-f \over 7900} \right]}~
    {\left[ cos^2 \left( \pi~f \over 3950 \right) + 0.1436 \right]}
\label{deadtime}
\end{equation}

\begin{equation}
    A~=~4~\left[ r_{in} t_d \right]~e^{- r_{in} t_d }~\approx~4~t_d r_o 
\label{Adead}
\end{equation}

\begin{equation}
    B~=~2~{\left[ r_{o} t_{vle} \right]}{\left[ r_{vle} t_{vle} \right]}
     ~\approx~{1 \over 2} A~\left[ t_{vle} \over t_d \right] t_{vle} r_{vle}
\label{Bvle}
\end{equation}

In Figure~\ref{figgx17+2}
we show the power density spectrum of GX 17+2 over a frequency range
that extends to $\sim50~$kHz (\cite{morgan99}).
The curve shows the Leahy normalized
power density on a linear scale near the expected value of 2 
as a function of the logarithm of the frequency.
The value of the power density falls below 2 for frequencies greater
than $\sim100~$Hz, clearly indicating that large deadtime corrections
are necessary to determine the intrinsic component of the power density.
The solid curve is the deadtime component of the
power density spectrum which approximately matches above $\sim2000~$Hz
as defined by equation~\ref{deadtime}
with values of A and B equal to 0.0597 and
0.0101 respectively. The value of C is consistent with 0 indicating
the low level of any background effects for source as bright as GX 17+2.
The match of the deadtime model to the
observed data is excellent above $\sim2000~$Hz as indicated by the
one sigma error bars (dotted stairstep curves).
In particular the broad dip in the power density near $\sim5000~$Hz
is fully explained as a deadtime correction due to VLE effects.
The lower dashed curve shows the deadtime model if B$~=~0$ with no VLE
correction. The upper dashed curve shows the deadtime model if the
VLE correction is arbitrarily increased by a factor of six.
This extreme form of the deadtime model shows a feature at
$\sim9~$kHz which is clearly in excess of the power density level
near that frequency. The correct deadtime model (solid curve) matches the
slight ``bump'' in the power density near $\sim9~$kHz. 
Figure~\ref{figgx17+2X} is an expanded version
of the data shown in figure~\ref{figgx17+2}.
The purpose of this figure is to clearly show the match of the
deadtime model to the slight bump in the power density near $\sim9~$kHz.
This feature corresponds to the second peak of the
$\left[ sin(x) \over x \right]^{2}$
function for the VLE correction to the deadtime model and therefore depends
on the ground calibrated value of $t_{vle}$.
The best values of A and B are approximately consistent with values of
$r_{o}~\sim~1500s^{-1}$ and $r_{vle}~\sim~120~s^{-1}$
for one of the five PCA modules. The implied total rates in the PCA
are approximately consistent with the actual measured rates.
The most conservative procedure for using this deadtime model is to allow the
values of A and B to vary in order to match the high frequency
component of the power density spectrum which is likely to be dominated by
deadtime effects. One can then derive approximate
values of $r_o$ and $r_{vle}$ by inverting equations ~\ref{Adead} and ~\ref{Bvle}
and then checking that these values are a rough
match to the typical measured rates.
With this interpretation of the deadtime correction Figure~\ref{figgx17+2}
shows the presence of the kHz QPO at $\sim600~$Hz
previously discovered by \cite{wijnands97}
but no significant intrinsic power above $\sim1000~$Hz.
This deadtime corrected interpretation of the power density spectrum
of GX 17+2 shows for the first time that there is very little
continuum power on either side of the kHz QPO feature. In summary,
this power density spectrum of GX 17+2 indicates that a simple deadtime model
with two free parameters A and B successfully explains the frequency
dependent form of the deadtime correction to an accuracy of 
$\sim1\%$ of the depression of the Leahy normalization
2 which is expected for pure Poisson noise with no deadtime correction.

Since our goal is to demonstrate that
Cen X-3 shows excess power density in the range from $300-2000~$Hz
above the level of the deadtime correction, we seek an independent source
of sufficient counting rate other that GX 17+2
which has no known intrinsic power in this frequency band.
Figure~\ref{figcygx1} shows
the power density spectrum of Cyg X-1 over a frequency range
that extends to $\sim6000~$Hz.
The data for the curve shown in Figure~\ref{figcygx1} were provided by
\cite{cui99} and show the Leahy normalized
power density on a linear scale near the expected value of 2
as a function of the logarithm of the frequency.
The value of the power density falls below 2 for frequencies greater
than $\sim100~$Hz clearly indicating that deadtime corrections
are necessary to determine the intrinsic component of the power density.
The solid curve is an approximate fit of the deadtime corrected
power density above $\sim300~$Hz as given by equation~\ref{deadtime}.
The approximate best values of A and B are 0.05 and 0.01 respectively
which corresponds to mean counting rate $r_o~\sim~1250~s^{-1}$ and
$r_{vle}~\sim~140~s^{-1}$ for one of the five PCA modules.
The implied total rates in the PCA are approximately consistent with
the actual measured rates and are similar to the rates for GX 17+2.
Again the background term (parameter C) is not needed because of the
brightness of CYG X-1.
The match of the deadtime model to the
observed data is excellent above $\sim300~$Hz. The lower
dashed curve shows the deadtime model if B$~=~0$ with no VLE
correction. The upper dashed curve shows the deadtime model if the
VLE correction is arbitrarily increased by a factor of three which
is clearly inconsistent with the power density spectrum.
The deadtime corrected interpretation of this spectrum indicates the
significant detection of intrinsic power from Cyg X-1 up to $\sim280~$Hz.
This interpretation extends and is consistent with prior published
observations of Cyg X-1 (\cite{cui97} and \cite{nowak99}).
For frequencies greater than $\sim300~$Hz the spectrum is consistent
with the deadtime model including the slight drop in power near $\sim2000~$Hz.
Since Cyg X-1 is a few times brighter than Cen X-3, these deadtime corrections
will be greater than those for Cen X-3. The results for the Cyg X-1 data
show that deadtime effects from $300-4000~$Hz are smooth and
correctable  with no unexplained features.
These data from Cyg X-1 were obtained in the similar PCA mode (SB\_125us\_0\_17\_1s))
as the data from Cen X-3 reported in this paper (SB\_62us\_0\_17\_1s).
In particular the energy range for the two data sets is identical.
One could successfully use a scaled version of the spectrum of Cyg X-1
between $300-4000~$Hz as the deadtime model for the Cen X-3 observations.

The stair step solid curve in
Figure~\ref{figleahy} shows
the power density of data from Cen X-3 over a frequency range
$\sim50~$Hz to $\sim10~$kHz. 
The stair step dotted curves above and below are the one
sigma errors of the measured power density.
These data are the same as shown in both
figures~\ref{figdata} and ~\ref{figfit} and are plotted here to clearly show the
level of the necessary deadtime correction.

The curve shows the Leahy normalized
power density on a linear scale near the expected value of 2
as a function of the logarithm of the frequency.
The value of the power density falls below 2 for frequencies greater
than $\sim50~$Hz clearly indicating that deadtime corrections
are necessary to determine the intrinsic component of the power density.
The upper solid curve is the best fit model (15 parameters) to the total
power density as a function of frequency.
The range and best fit values of each of the 15 parameters is shown in Table 2.
Three of the parameters are the components A, B and C of the deadtime model
in the form given by equations ~\ref{deadtime}, ~\ref{Adead} and ~\ref{Bvle}. 
The lower solid curve shows the deadtime model for the best fit
values of A, B and C (see Table 2).
These data for Cen X-3 were obtained with a setting for $t_{vle}$ of
$\sim61\mu s~$(level 1)
in contrast a higher setting of $142\mu s$(level 2) for the GX 17+2 and Cyg X-1 data.
The lower setting decreases any VLE deadtime effect and was set to this lower
value for the Cen X-3 observations precisely to decrease any VLE effect.
The values of A and B are used to estimate typical values for
$r_o$ ($\sim310~s^{-1}$) and
$r_{vle}$ ($\sim40~s^{-1}$)
which roughly agree with the observed average rates.
The match of the deadtime model to the
observed data is excellent above $\sim2~$kHz. The lower
dashed curve shows the deadtime model if B$~=~0$ with no VLE
correction. The upper dashed curve shows the deadtime model if the
VLE correction is arbitrarily increased by a factor of six.
Such a model roughly matches the level of the power density in
the frequency range from $100-1000~$Hz but is completely
inconsistent with the power density from $1-10~$kHz.
Clearly the excess power density in the frequency range $100-2000~$Hz
can not be explained as a VLE deadtime effect.
The dotted continuous curves just above and below the lower solid
curve show the affect on the deadtime model if the VLE correction
factor (parameter B) is varied from a low to a high value
corresponding to the $50\%$ confidence range.
A correct  VLE deadtime correction does effect
the determination of the intrinsic level of the
continuum power in the $100-1000~$Hz range. 
However the continuum component (see dashed-dotted curve) is well above the
level of the deadtime component (lower solid curve).
The lower short-dash-log-dash curve is the upper limit to the background
as determine by the free parameter C. It shows a drop at 2000 Hz
followed by a small peak at 4000 Hz which reachs but does not exceed the
measured power density at 4000 Hz. This upper limit for C greatly exceeds
any expected level of the background which is likely at least full
order of magitude lower. Even so, a value of C that is increased by a factor
8 (see the upper short-dash-long-dash curve) does roughly match
the overall shape of the power density spectrum from 100-2000 Hz.
This artificially amplified value of C would create a strong peak at
4000 Hz which is well above the measured level of the power density
near 4000 Hz. Estimates for the contribution of the background are also
shown in Figure~\ref{figdata}. An additive background contribution is not a
significant component of the power density spectrum of Cen X-3.
\clearpage

%\input{bib.tex}

%\input{figure.tex}

% Psfig/TeX Release 1.2
% dvi2ps-li version
%
% All software, documentation, and related files in this distribution of
% psfig/tex are Copyright 1987, 1988 Trevor J. Darrell
%
% Permission is granted for use and non-profit distribution of psfig/tex 
% providing that this notice be clearly maintained, but the right to
% distribute any portion of psfig/tex for profit or as part of any commercial
% product is specifically reserved for the author.
%
% $Header: psfig.tex,v 1.9 88/01/08 17:42:01 trevor Exp $
% $Source: $
%
% Thanks to Greg Hager (GDH) and Ned Batchelder for their contributions
% to this project.
%
\catcode`\@=11\relax
\newwrite\@unused
\def\typeout#1{{\let\protect\string\immediate\write\@unused{#1}}}
\typeout{psfig/tex 1.2-dvi2ps-li}

%% Here's how you define your figure path.  Should be set up with null
%% default and a user useable definition.

\def\figurepath{./}
\def\psfigurepath#1{\edef\figurepath{#1}}

%
% @psdo control structure -- similar to Latex @for.
% I redefined these with different names so that psfig can
% be used with TeX as well as LaTeX, and so that it will not 
% be vunerable to future changes in LaTeX's internal
% control structure,
%
\def\@nnil{\@nil}
\def\@empty{}
\def\@psdonoop#1\@@#2#3{}
\def\@psdo#1:=#2\do#3{\edef\@psdotmp{#2}\ifx\@psdotmp\@empty \else
    \expandafter\@psdoloop#2,\@nil,\@nil\@@#1{#3}\fi}
\def\@psdoloop#1,#2,#3\@@#4#5{\def#4{#1}\ifx #4\@nnil \else
       #5\def#4{#2}\ifx #4\@nnil \else#5\@ipsdoloop #3\@@#4{#5}\fi\fi}
\def\@ipsdoloop#1,#2\@@#3#4{\def#3{#1}\ifx #3\@nnil 
       \let\@nextwhile=\@psdonoop \else
      #4\relax\let\@nextwhile=\@ipsdoloop\fi\@nextwhile#2\@@#3{#4}}
\def\@tpsdo#1:=#2\do#3{\xdef\@psdotmp{#2}\ifx\@psdotmp\@empty \else
    \@tpsdoloop#2\@nil\@nil\@@#1{#3}\fi}
\def\@tpsdoloop#1#2\@@#3#4{\def#3{#1}\ifx #3\@nnil 
       \let\@nextwhile=\@psdonoop \else
      #4\relax\let\@nextwhile=\@tpsdoloop\fi\@nextwhile#2\@@#3{#4}}
\def\psdraft{
	\def\@psdraft{0}
	%\typeout{draft level now is \@psdraft \space . }
}
\def\psfull{
	\def\@psdraft{100}
	%\typeout{draft level now is \@psdraft \space . }
}
\psfull
\newif\if@prologfile
\newif\if@postlogfile
\newif\if@noisy
\def\pssilent{
	\@noisyfalse
}
\def\psnoisy{
	\@noisytrue
}
\psnoisy
%%% These are for the option list.
%%% A specification of the form a = b maps to calling \@p@@sa{b}
\newif\if@bbllx
\newif\if@bblly
\newif\if@bburx
\newif\if@bbury
\newif\if@height
\newif\if@width
\newif\if@rheight
\newif\if@rwidth
\newif\if@clip
\newif\if@verbose
\def\@p@@sclip#1{\@cliptrue}

%%% GDH 7/26/87 -- changed so that it first looks in the local directory,
%%% then in a specified global directory for the ps file.

\def\@p@@sfile#1{\def\@p@sfile{null}%
	        \openin1=#1
		\ifeof1\closein1%
		       \openin1=\figurepath#1
			\ifeof1\typeout{Error, File #1 not found}
			\else\closein1
			    \edef\@p@sfile{\figurepath#1}%
                        \fi%
		 \else\closein1%
		       \def\@p@sfile{#1}%
		 \fi}
\def\@p@@sfigure#1{\def\@p@sfile{null}%
	        \openin1=#1
		\ifeof1\closein1%
		       \openin1=\figurepath#1
			\ifeof1\typeout{Error, File #1 not found}
			\else\closein1
			    \def\@p@sfile{\figurepath#1}%
                        \fi%
		 \else\closein1%
		       \def\@p@sfile{#1}%
		 \fi}

\def\@p@@sbbllx#1{
		%\typeout{bbllx is #1}
		\@bbllxtrue
		\dimen100=#1
		\edef\@p@sbbllx{\number\dimen100}
}
\def\@p@@sbblly#1{
		%\typeout{bblly is #1}
		\@bbllytrue
		\dimen100=#1
		\edef\@p@sbblly{\number\dimen100}
}
\def\@p@@sbburx#1{
		%\typeout{bburx is #1}
		\@bburxtrue
		\dimen100=#1
		\edef\@p@sbburx{\number\dimen100}
}
\def\@p@@sbbury#1{
		%\typeout{bbury is #1}
		\@bburytrue
		\dimen100=#1
		\edef\@p@sbbury{\number\dimen100}
}
\def\@p@@sheight#1{
		\@heighttrue
		\dimen100=#1
   		\edef\@p@sheight{\number\dimen100}
		%\typeout{Height is \@p@sheight}
}
\def\@p@@swidth#1{
		%\typeout{Width is #1}
		\@widthtrue
		\dimen100=#1
		\edef\@p@swidth{\number\dimen100}
}
\def\@p@@srheight#1{
		%\typeout{Reserved height is #1}
		\@rheighttrue
		\dimen100=#1
		\edef\@p@srheight{\number\dimen100}
}
\def\@p@@srwidth#1{
		%\typeout{Reserved width is #1}
		\@rwidthtrue
		\dimen100=#1
		\edef\@p@srwidth{\number\dimen100}
}
\def\@p@@ssilent#1{ 
		\@verbosefalse
}
\def\@p@@sprolog#1{\@prologfiletrue\def\@prologfileval{#1}}
\def\@p@@spostlog#1{\@postlogfiletrue\def\@postlogfileval{#1}}
\def\@cs@name#1{\csname #1\endcsname}
\def\@setparms#1=#2,{\@cs@name{@p@@s#1}{#2}}
%
% initialize the defaults (size the size of the figure)
%
\def\ps@init@parms{
		\@bbllxfalse \@bbllyfalse
		\@bburxfalse \@bburyfalse
		\@heightfalse \@widthfalse
		\@rheightfalse \@rwidthfalse
		\def\@p@sbbllx{}\def\@p@sbblly{}
		\def\@p@sbburx{}\def\@p@sbbury{}
		\def\@p@sheight{}\def\@p@swidth{}
		\def\@p@srheight{}\def\@p@srwidth{}
		\def\@p@sfile{}
		\def\@p@scost{10}
		\def\@sc{}
		\@prologfilefalse
		\@postlogfilefalse
		\@clipfalse
		\if@noisy
			\@verbosetrue
		\else
			\@verbosefalse
		\fi

}
%
% Go through the options setting things up.
%
\def\parse@ps@parms#1{
	 	\@psdo\@psfiga:=#1\do
		   {\expandafter\@setparms\@psfiga,}}
%
% Compute bb height and width
%
\newif\ifno@bb
\newif\ifnot@eof
\newread\ps@stream
\def\bb@missing{
	\if@verbose{
		\typeout{psfig: searching \@p@sfile \space  for bounding box}
	}\fi
	\openin\ps@stream=\@p@sfile
	\no@bbtrue
	\not@eoftrue
	\catcode`\%=12
	\loop
		\read\ps@stream to \line@in
		\global\toks200=\expandafter{\line@in}
		\ifeof\ps@stream \not@eoffalse \fi
		%\typeout{ looking at :: \the\toks200 }
		\@bbtest{\toks200}
		\if@bbmatch\not@eoffalse\expandafter\bb@cull\the\toks200\fi
	\ifnot@eof \repeat
	\catcode`\%=14
}	
\catcode`\%=12
\newif\if@bbmatch
\def\@bbtest#1{\expandafter\@a@\the#1%%BoundingBox:\@bbtest\@a@}
\long\def\@a@#1%%BoundingBox:#2#3\@a@{\ifx\@bbtest#2\@bbmatchfalse\else\@bbmatchtrue\fi}
\long\def\bb@cull#1 #2 #3 #4 #5 {
	\dimen100=#2 bp\edef\@p@sbbllx{\number\dimen100}
	\dimen100=#3 bp\edef\@p@sbblly{\number\dimen100}
	\dimen100=#4 bp\edef\@p@sbburx{\number\dimen100}
	\dimen100=#5 bp\edef\@p@sbbury{\number\dimen100}
	\no@bbfalse
}
\catcode`\%=14
\def\compute@bb{
		\no@bbfalse
		\if@bbllx \else \no@bbtrue \fi
		\if@bblly \else \no@bbtrue \fi
		\if@bburx \else \no@bbtrue \fi
		\if@bbury \else \no@bbtrue \fi
		\ifno@bb \bb@missing \fi
		\ifno@bb \typeout{FATAL ERROR: no bb supplied or found}
			\no-bb-error
		\fi
		\count203=\@p@sbburx
		\count204=\@p@sbbury
		\advance\count203 by -\@p@sbbllx
		\advance\count204 by -\@p@sbblly
		\edef\@bbw{\number\count203}
		\edef\@bbh{\number\count204}
		%\typeout{ bbh = \@bbh, bbw = \@bbw }
}
%
% \in@hundreds performs #1 * (#2 / #3) correct to the hundreds,
%	then leaves the result in @result
%
\def\in@hundreds#1#2#3{\count240=#2 \count241=#3
		     \count100=\count240	% 100 is first digit #2/#3
		     \divide\count100 by \count241
		     \count101=\count100
		     \multiply\count101 by \count241
		     \advance\count240 by -\count101
		     \multiply\count240 by 10
		     \count101=\count240	%101 is second digit of #2/#3
		     \divide\count101 by \count241
		     \count102=\count101
		     \multiply\count102 by \count241
		     \advance\count240 by -\count102
		     \multiply\count240 by 10
		     \count102=\count240	% 102 is the third digit
		     \divide\count102 by \count241
		     \count200=#1\count205=0
		     \count201=\count200
			\multiply\count201 by \count100
		 	\advance\count205 by \count201
		     \count201=\count200
			\divide\count201 by 10
			\multiply\count201 by \count101
			\advance\count205 by \count201
		     \count201=\count200
			\divide\count201 by 100
			\multiply\count201 by \count102
			\advance\count205 by \count201
		     \edef\@result{\number\count205}
}
\def\compute@wfromh{
		% computing : width = height * (bbw / bbh)
		\in@hundreds{\@p@sheight}{\@bbw}{\@bbh}
		%\typeout{ \@p@sheight * \@bbw / \@bbh, = \@result }
		\edef\@p@swidth{\@result}
		%\typeout{w from h: width is \@p@swidth}
}
\def\compute@hfromw{
		% computing : height = width * (bbh / bbw)
		\in@hundreds{\@p@swidth}{\@bbh}{\@bbw}
		%\typeout{ \@p@swidth * \@bbh / \@bbw = \@result }
		\edef\@p@sheight{\@result}
		%\typeout{h from w : height is \@p@sheight}
}
\def\compute@handw{
		\if@height 
			\if@width
			\else
				\compute@wfromh
			\fi
		\else 
			\if@width
				\compute@hfromw
			\else
				\edef\@p@sheight{\@bbh}
				\edef\@p@swidth{\@bbw}
			\fi
		\fi
}
\def\compute@resv{
		\if@rheight \else \edef\@p@srheight{\@p@sheight} \fi
		\if@rwidth \else \edef\@p@srwidth{\@p@swidth} \fi
}
%		
% Compute any missing values
\def\compute@sizes{
	\compute@bb
	\compute@handw
	\compute@resv
}
%
% \psfig
% usage : \psfig{file=, height=, width=, bbllx=, bblly=, bburx=, bbury=,
%			rheight=, rwidth=, clip=}
%
% "clip=" is a switch and takes no value, but the `=' must be present.
\def\psfig#1{\vbox {
	% do a zero width hard space so that a single
	% \psfig in a centering enviornment will behave nicely
	%{\setbox0=\hbox{\ }\ \hskip-\wd0}
	%
	\ps@init@parms
	\parse@ps@parms{#1}
	\compute@sizes
	\ifnum\@p@scost<\@psdraft{
		\if@verbose{
			\typeout{psfig: including \@p@sfile \space }
		}\fi
		\special{ pstext="\@p@swidth \space 
			\@p@sheight \space
			\@p@sbbllx \space \@p@sbblly \space 
			\@p@sbburx  \space 
			\@p@sbbury \space startTexFig" \space}
		\if@clip{
			\if@verbose{
				\typeout{(clip)}
			}\fi
			\special{ pstext="doclip \space"}
		}\fi
		\if@prologfile
		    \includegraphics{\@prologfileval} \fi
		\includegraphics{\@p@sfile}
		\if@postlogfile
		    \includegraphics{\@postlogfileval} \fi
		\special{pstext=endTexFig \space }
		% Create the vbox to reserve the space for the figure
		\vbox to \@p@srheight true sp{
			\hbox to \@p@srwidth true sp{
				\hss
			}
			\vss
		}
	}\else{
		% draft figure, just reserve the space and print the
		% path name.
		\vbox to \@p@srheight true sp{
		\vss
			\hbox to \@p@srwidth true sp{
				\hss
				\if@verbose{
					\@p@sfile
				}\fi
				\hss
			}
		\vss
		}
	}\fi
}}
\def\psglobal{\typeout{psfig: PSGLOBAL is OBSOLETE; use psprint -m instead}}
\catcode`\@=12\relax

\epsfysize 5.0in

\begin{figure}
%\special{psfile=fig1.eps}
%\plotone{fig1.eps}
\centerline{\psfig{figure=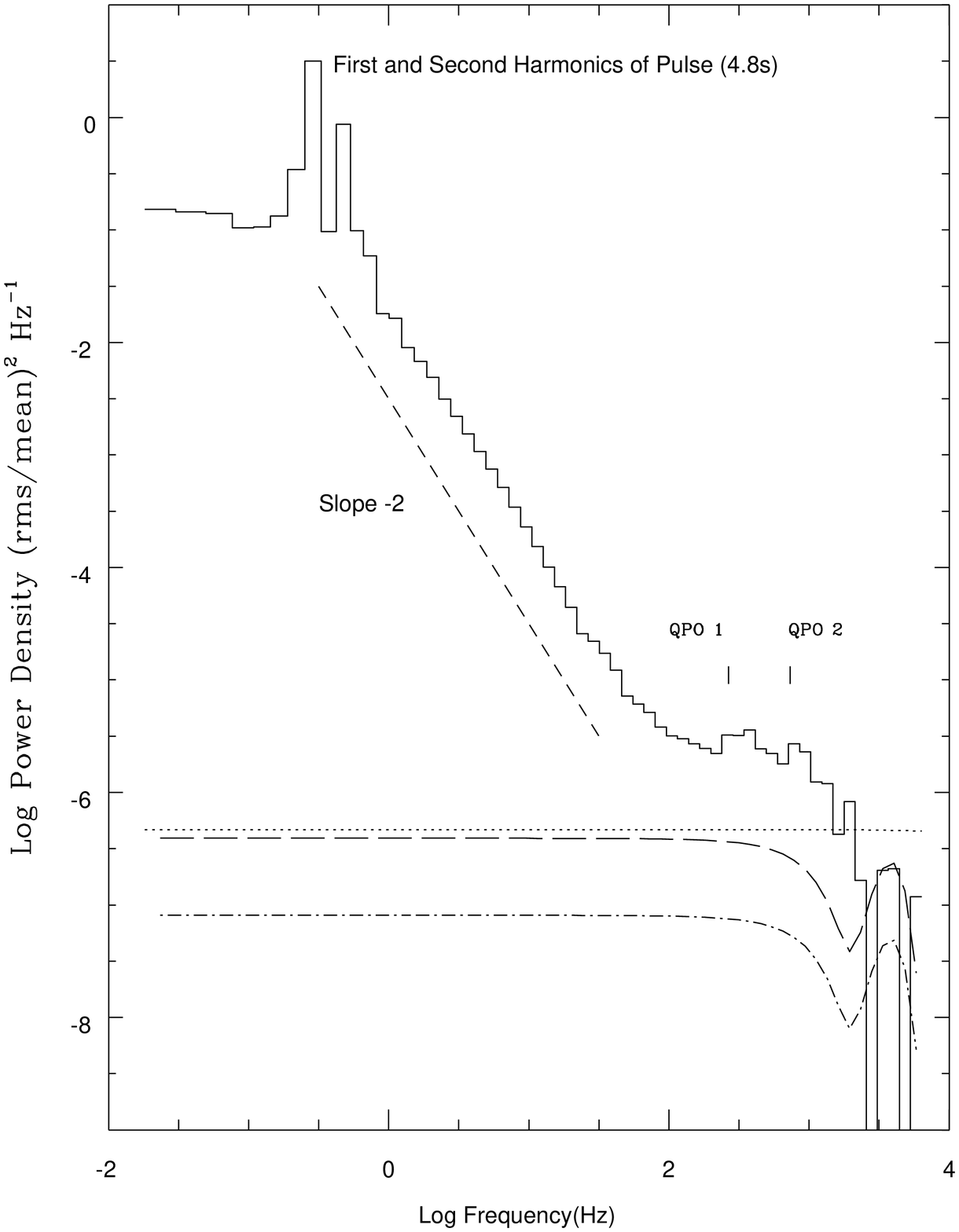,height=3.2in}}
%\begin{center}
%\end{center}
\caption{ The curve is the observed average power density in units
$\log_{10} \left[ \left( {\rm rms \over mean} \right){\rm Hz}^{-1}
\right]$ as a function of $\log_{10} \left[{\rm frequency~(Hz)}
\right]$. Note the presence of the first and second harmonics of
the pulse (0.21 and 0.42 Hz) due to the rotation of the neutron
star, the power law continuum below $\sim100~$Hz with a slope of
$\sim-2$ (dashed line), the flattening of the curve near $100~$Hz,
the two kHz QPO peaks, and the continuum power between
$100-2500~$Hz.
This intrinsic power density spectrum 
has been corrected for the effects of Poisson noise
including the slight frequency dependence caused by
deadtime effects of the PCA (see appendix A).
The dotted curve shows an estimate of the upper limit to the
level of the deadtime corrected Poisson level. 
Power density at high frequency that falls below the dotted line
is not confirmed as intrinsic to Cen X-3.
The long-dashed curve just below the dotted curve shows the upper limit
to any additive contribution from the background. This non-source component
of the power density spectrum can not be much larger that shown here
otherwise the peak at 4000 Hz would be detectable. The lower 
dot-dashed curve is the most likely level of the background based on the
correction procedure (see Appendix A). The actual level of the background
is likely another order of magitude smaller than the dot-dashed curve. \label{figdata}}
\end{figure}

\begin{figure}
%\special{psfile=fig2.eps}
%\plotone{fig2.eps}
\centerline{\psfig{figure=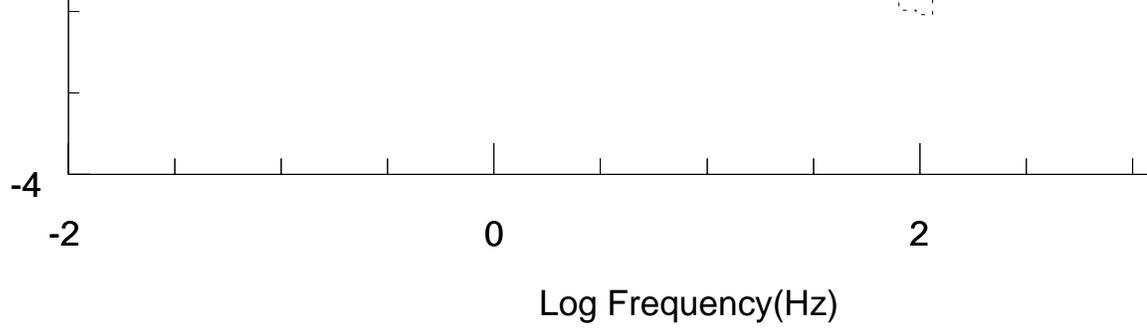,height=5.0in}}
%\begin{center}
%\end{center}
\caption{The curve of measured data (solid stair step) is the same
as in Figure~\ref{figdata}, but has now been multiplied by
frequency. Dotted lines are plotted above and below the data at $1
\sigma$ error levels. Dashed lines are plotted horizontally at
levels of 2, 5 and 100 $\% \left({\rm rms \over mean} \right)$.
Note that the pulsed fraction is at a level of several tens of
$\%$. The two kHz QPO peaks and the continuum power above $100~$Hz
have a total power bounded between 2 and 5$\%$ rms. The best fit
to the data plotted as a continuous solid curve tracks the
measurements. \label{figfit}}
\end{figure}

\begin{figure}
%\special{psfile=fig3.eps}
%\plotonefig3.eps}
\centerline{\psfig{figure=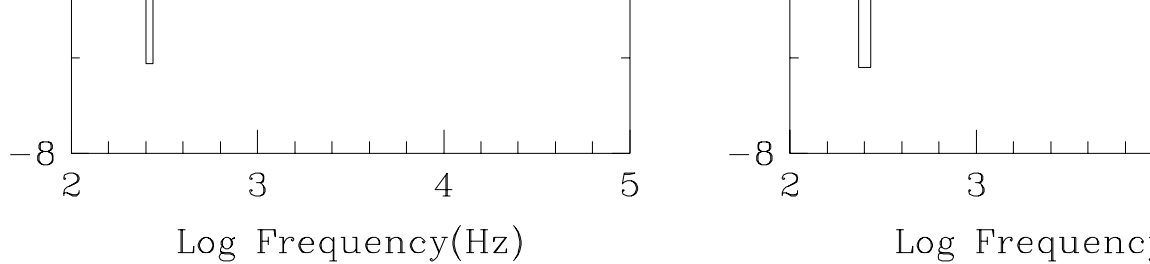,height=5.0in}}
%\begin{center}
%\end{center}

\caption{The left panel is a log-log plot of power density spectra
in units
%$\log_{10}{\rm \left[ Power Density \left( rms \over mean \right) \right]}$ of models A,B and C versus $\log_{10} {\rm \left[ Frequency(Hz) \right]}$.
The right panel is a smoothed
representation of the data in the left panel and is more
appropriate for direct comparison with the observed data. The
vertical scale has been shifted up by 2 and 4 decades for models B
and C respectively for illustrative purposes. Note that all three
models show evidence of PBOs and PB continuum power.
Lines that indicate power laws of the form $f^{-2}$ (dotted) and
$f^{-\left( 5 \over 3 \right)}$ (dashed) are indicated in the right panel. These
slopes at high frequency are roughly consistent with the results
that we reported previously
i%(\cite{klein96b}).
\label{figmodel}
}

\end{figure}

\begin{figure}
%\special{psfile=fig4.eps}
%\plotone{fig4.eps}
\centerline{\psfig{figure=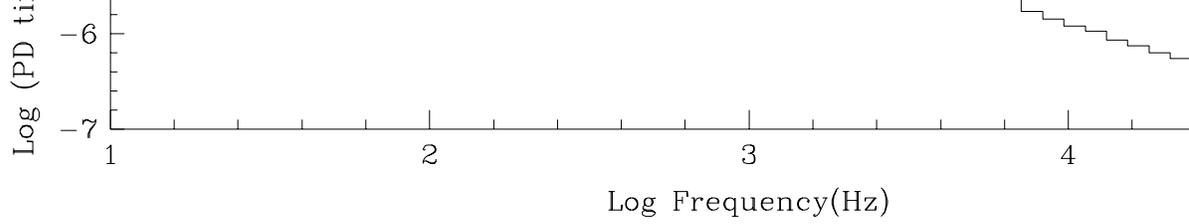,height=5.0in}}
%\begin{center}
%\end{center}
\caption{The upper curve in each panel is the observed power
density of Cen X-3 from Figure~\ref{figfit}. The error bars are
shown to facilitate a quantitative comparison to the theoretical models.
Beyond $\sim1800~$Hz the $2\sigma$ upper limit is shown as a dotted curve.
Also the low frequency power law fitted to the observed data has been
subtracted since this component of the observed spectrum is not
included in the physics incorporated into the theoretical models.
The power density spectra of the three models A, B and C
are shown on the same absolute scale. These are the smoothed
versions of the spectra appropriate for direct comparison to the
observed spectrum of Cen X-3. The three models have been correctly
red shifted by the appropriate amount for the surface of a neutron
star. \label{figcompare}}
\end{figure}

\begin{figure}
\centerline{\psfig{figure=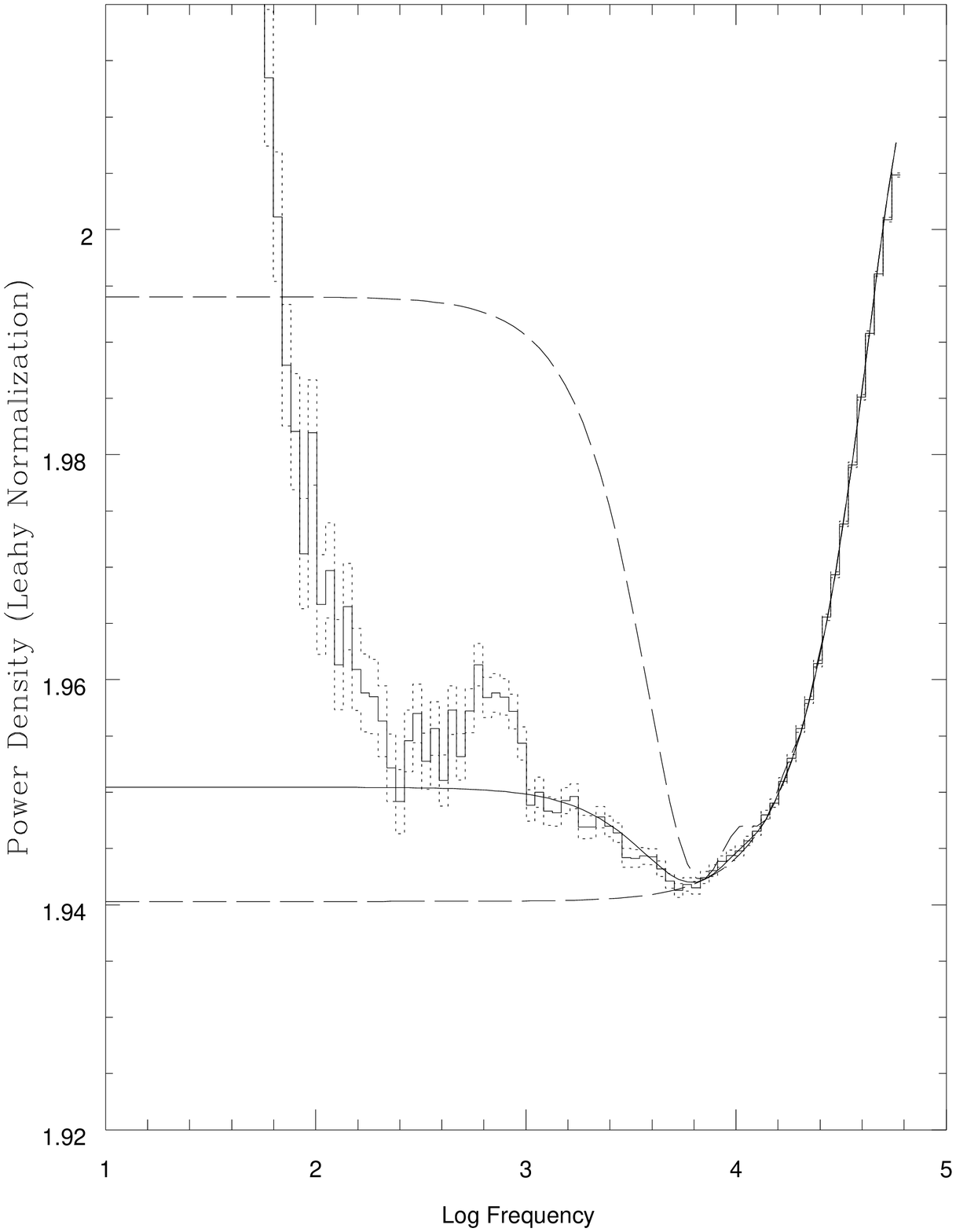,height=3.0in}}
%\begin{center}
%\end{center}

\caption{
The power density of data from GX 17+2 over a frequency range
that extends to $\sim50~$kHz. The curve shows the Leahy normalized
power density on a linear scale near the expected value of 2 
as a function of the logarithm of the frequency.
The value of the power density falls below 2 for frequencies greater
than $\sim100~$Hz clearly indicating that large deadtime corrections
are necessary to determine the intrinsic component of the power density.
The solid curve is an approximate fit of the deadtime corrected
power density above $\sim2000~$Hz as given by equations~\ref{deadtime},
~\ref{Adead} and ~\ref{Bvle}. The match of the deadtime model to the
observed data is excellent above $\sim2000~$Hz. The lower
dashed curve shows the deadtime model if B$~=~0$ with no VLE
correction. The upper dashed curve shows the deadtime model if the
VLE correction is arbitrarily increased by a factor of six.
This extreme form of the deadtime model shows a feature at
$\sim9~$kHz which is clearly in excess of the power density level
of the data. The correct deadtime model (solid curve) matches the
slight "bump" in the power density near $\sim9~$kHz.
\label{figgx17+2}
}

\end{figure}

\begin{figure}
\centerline{\psfig{figure=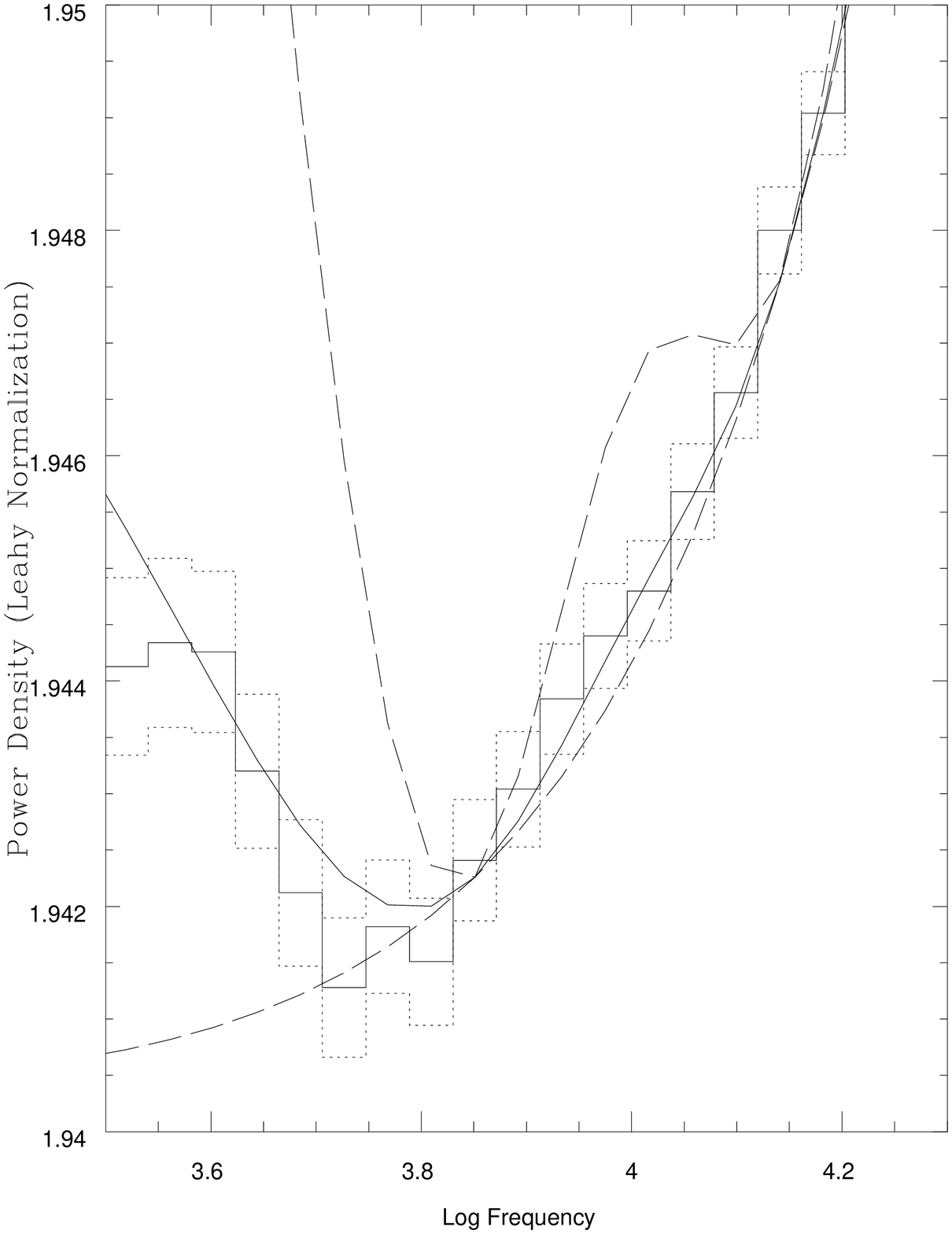,height=4.0in}}
%\begin{center}
%\end{center}
\caption{
This figure is an expanded version of the data shown in figure~\ref{figgx17+2}.
The purpose of this figure is to clearly show the match of the
deadtime model to the slight bump in the power density at $\sim9~$kHz.
This feature corresponds to the second peak of the $\left[ sin(x) \over x \right]^2$
function for the VLE correction to the deadtime model. \label{figgx17+2X} }
\end{figure}

\begin{figure}
\centerline{\psfig{figure=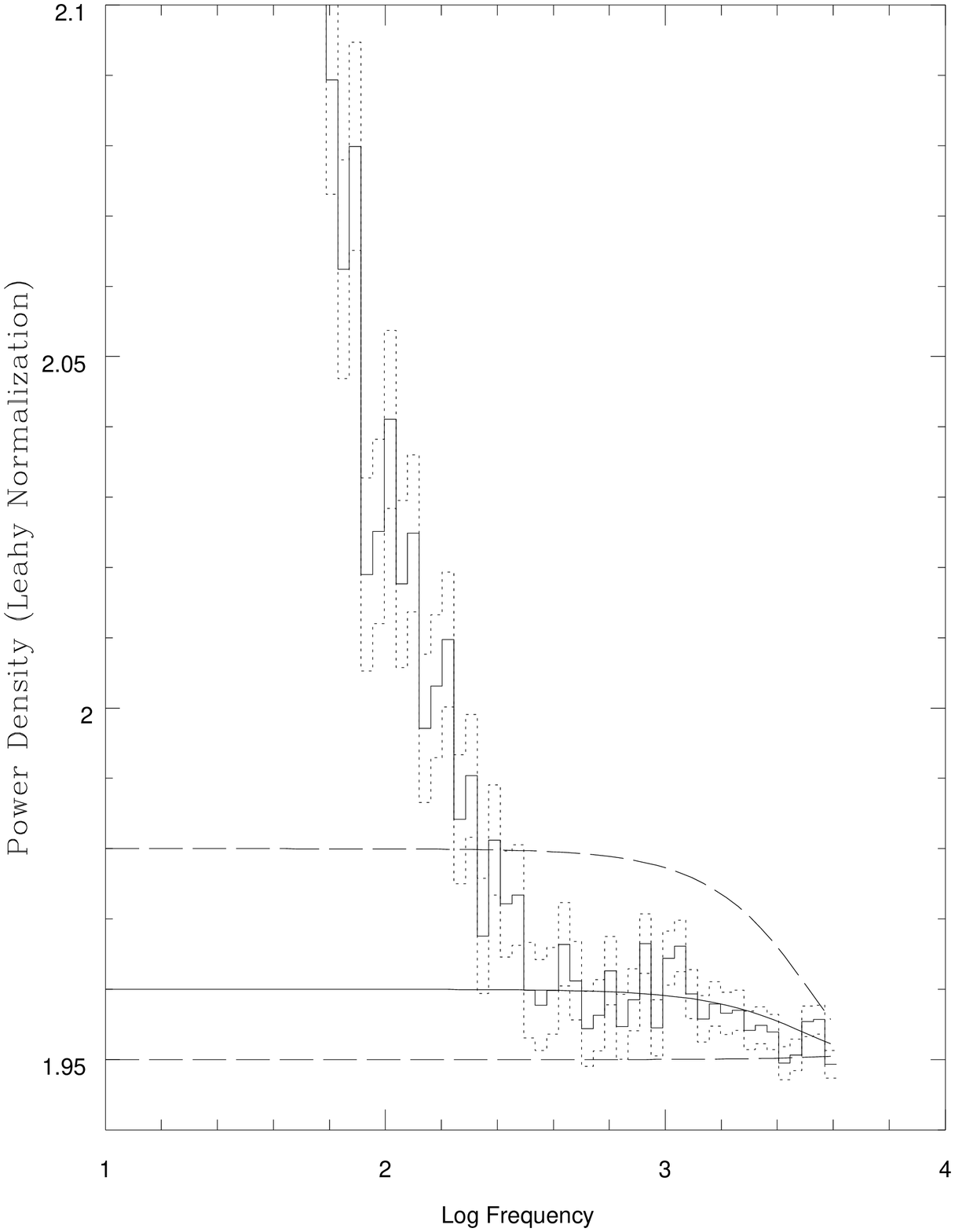,height=4.0in}}
%\begin{center}
%\end{center}
\caption{
The power density of data from Cyg X-1 over a frequency range
that extends to $\sim6000~$Hz. The curve shows the Leahy normalized
power density on a linear scale near the expected value of 2 at
as a function of the logarithm of the frequency.
The value of the power density falls below 2 for frequencies greater
than $\sim100~$Hz clearly indicating that deadtime corrections
are necessary to determine the intrinsic component of the power density.
The solid curve is an approximate fit of the deadtime corrected
power density above $\sim300~$Hz as given by equations~\ref{deadtime},
~\ref{Adead} and ~\ref{Bvle}. The match of the deadtime model to the
observed data is excellent above $\sim300~$Hz. The lower
dashed curve shows the deadtime model if B$~=~0$ with no VLE
correction. The upper dashed curve shows the deadtime model if the
VLE correction is arbitrarily increased
by a factor of three. \label{figcygx1}}
\end{figure}

\begin{figure}
\centerline{\psfig{figure=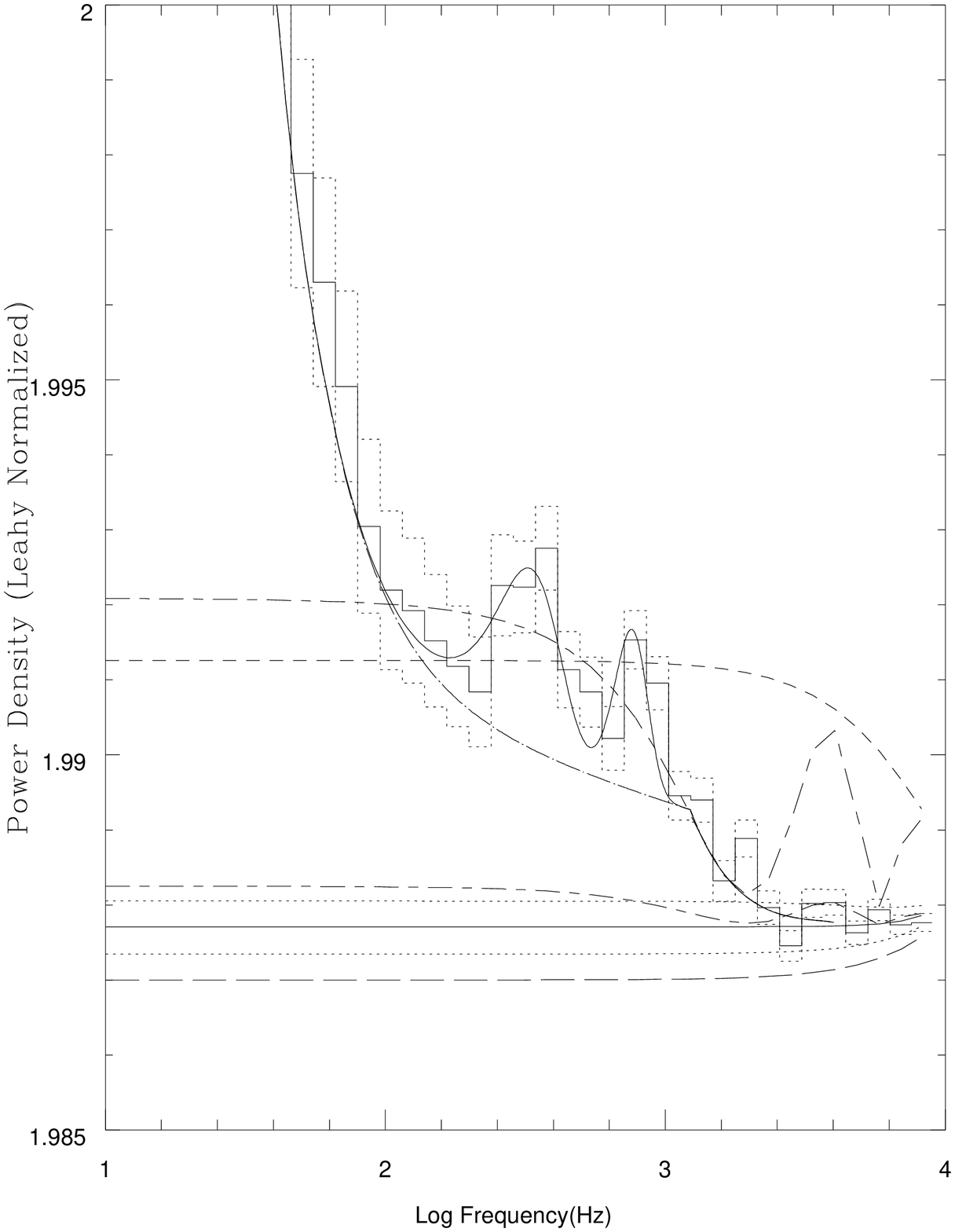,height=3.0in}}
%\begin{center}
%\end{center}
\caption{
The power density of data from Cen X-3 over a frequency range
$\sim50~$Hz to $\sim10~$kHz is the stair step solid curve.
The stair step dotted curves above and below are the one
sigma errors of the measured power density.
These data are the same as shown in both
figures~\ref{figdata} and ~\ref{figfit} and are plotted to clearly show the
level of the necessary deadtime correction.
The curve shows the Leahy normalized
power density on a linear scale near the expected value of 2
as a function of the logarithm of the frequency.
The upper solid curve is the best fit model (15 parameters) to the total
power density as a function of frequency.
The lower solid curve is the best fit of the deadtime correction to the
power density. The match of the deadtime model to the
observed data is excellent above $\sim2~$kHz. The lower
dashed curve shows the deadtime model if B$~=~0$ with no VLE
correction. The upper dashed curve shows the deadtime model if the
VLE correction is arbitrarily increased by a factor of six.
The dotted continuous curves just above and below the lower solid
curve show the effect on the deadtime model if the VLE correction
factor (parameter B) is varied from a low to a high value that
corresponds to a 50$\%$ confidence range.
The continuum component (see dashed-dotted curve) is well above the
level of the deadtime component (lower solid curve). \label{figleahy}}
\end{figure}

\begin{figure}
\centerline{\psfig{figure=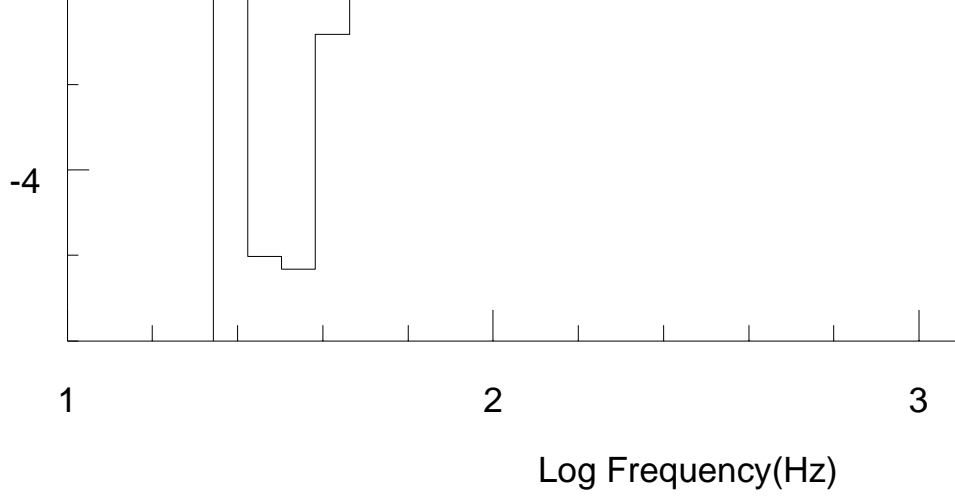,height=3.0in}}
%\begin{center}
%\end{center}
\caption{
The power density residuals of data from Cen X-3 over a frequency range
$\sim50~$Hz to $\sim10~$kHz is the stair step solid curve.
The model minus data residuals are plotted in units of the local measured one sigma error
for each data point. These model and data are the same as shown in figure~\ref{figleahy}.
The model approximately uniformly fits the data above $\sim100~$Hz and therefore all
parameters related to the kHz QPO and continuum are formally well defined.
The power law model below $\sim100~$Hz is clearly not sufficiently complex to
completely model the behavior of Cen X-3 at low frequency. This difficulty with the model is amplified
by the excellent signal to noise below $\sim100~$Hz. \label{figleahyerr}}
\end{figure}

\begin{deluxetable}{ccrrrcc}
\footnotesize \tablecaption{Parameters of Cen X-3}
\label{tblcenx3}
\tablewidth{0pt}
\tablehead{
\colhead{Parameter} &
\colhead{Variable} &
\colhead{Value} &
\colhead{Lower Value} &
\colhead{Upper Value} &
\colhead{Units} &
\colhead{Reference}
}
\startdata
Mass of NS       &  $M_{ns}$     &  1.4   &  1.3 &  1.5 & solar mass           & \cite{rappaport83} \nl
Radius of NS     &  $R_{ns}$     &  10.0  &  9.0 & 17.0 & km                   & \cite{baym79} \nl
Cyclotron Line   &  $E_{C}$      & 28.5   & 27.5 & 29.5 & keV                  & \cite{sax98} \nl
Magnetic Field   &  $B$          &  3.2   & 3.1  &  3.3 & $10^{12}$ Gauss      & \cite{sax98} \nl
Distance         &  $D$          & 10.0   &  9.0 & 11.0 & kpc                  & \cite{hutchings79} \nl
X-ray Luminosity &  $L_{x}$      &  9.4   &  9.0 &  9.8 & $10^{37} {\rm erg s}^{-1}$ & (this paper)\nl
Polar Cap Radius & $\theta_{c}$  &  0.25  &  0.1 &  0.4 & radians              & (this paper)\nl

\enddata

\end{deluxetable}

\begin{deluxetable}{ccrrrc}
\footnotesize
\tablecaption{Fitted Model Parameters}
\label{tbldata}
\tablewidth{0pt}
\tablehead{
\colhead{Parameter} &
\colhead{Variable} &
\colhead{Value} &
\colhead{Lower Value} &
\colhead{Upper Value} &
\colhead{Units}
}
\startdata
Low Corner Frequency  &  $F_{l}$  &  63     &  46     &  76    & Hz \nl
High Corner Frequency &  $F_{h}$  & 1228    &  923    & 1662   & Hz \nl
PD($F_{l}$)            &  $P_{l}$  & 2.50 & 2.05 &  2.95 & $10^{-6}\left({\rm rms \over mean}\right)^2{\rm Hz}^{-1}$ \nl
Low PD Slope          &  $S_{l}$  &  -2.15  &  -3.89  & -1.04  & dimensionless \nl
Mid PD Slope          &  $S_{m}$  &  -0.27  &  -0.37  & -0.18  & dimensionless \nl
High PD Slope         &  $S_{h}$  &  -2.61  &  -8.1   &  -1.5  & dimensionless \nl
Frequency of QPO 1    &  $F_{1}$  &  331    &   260   &  407   & Hz \nl
Amplitude of QPO 1    &  $A_{1}$  &  3.1    &  2.1    & 3.9    & $\%$ rms \nl
Equiv. width of QPO 1 &  $A_{1}$  &  712    &   329   & 1106   & Hz \nl
Width of QPO 1 ($1~\sigma$) &  $W_{1}$  &   89    &  48     & 213    & Hz \nl
Width of QPO 1 (FWHM)       &  $W_{1}$  &  214    & 115     & 511    & Hz \nl
Q of QPO 1 (FWHM)     & $F_{1}/W_{1}$   &  1.5    &  1.2    & 1.9    & dimensionless \nl
Frequency of QPO 2    &  $F_{2}$  &  761    &  671    &  849   & Hz \nl
Amplitude of QPO 2    &  $A_{2}$  &  3.1    &  2.0    &  4.0   & $\%$ rms \nl
Equiv. width of QPO 2 &  $A_{2}$  &  969    &  426    & 1534   & Hz \nl
Width of QPO 2 ($1~\sigma$) &  $W_{2}$  &  103    &  36     &  262   & Hz \nl
Width of QPO 2 (FWHM)       &  $W_{2}$  &  247    &  86     &  629   & Hz \nl
Q of QPO 2 (FWHM)     & $F_{2}/W_{2}$   &  3.1    &  2.7    &   3.4  & dimensionless \nl
Deadtime Factor       &    A      & 0.01304 & 0.01285 & 0.01328 & dimensionless \nl
VLE Deadtime Factor   &    B      & 0.00071 & 0.00048 & 0.00095 & dimensionless \nl
Background            &    C      & 0.00011 & -0.00032 & 0.00055   & dimensionless \nl
\enddata

\end{deluxetable}

\end{document}